\documentclass[aps,prd,10pt,twocolumn,notitlepage,superscriptaddress,nofootinbib,numbers,floatfix]{revtex4-2}

\usepackage[utf8]{inputenc}
\usepackage[T1]{fontenc}
\usepackage{lmodern}
\usepackage{microtype}
\usepackage{amsmath,amssymb,amsfonts,mathrsfs}
\usepackage{bm}
\usepackage{graphicx}
\usepackage[dvipsnames]{xcolor}
\usepackage{etoolbox}
\usepackage{hyperref}
\hypersetup{colorlinks=true,citecolor=black,linkcolor=black,urlcolor=black,breaklinks=true}
\newcommand{\revMagnetic}[1]{#1}
\usepackage[capitalize]{cleveref}
\usepackage{tikz}

\definecolor{lime}{HTML}{A6CE39}
\DeclareRobustCommand{\orcidicon}{%
	\begin{tikzpicture}
		\draw[lime, fill=lime] (0,0) 
		circle [radius=0.16] 
		node[white] {{\fontfamily{qag}\selectfont \tiny ID}};
		\draw[white, fill=white] (-0.0625,0.095) 
		circle [radius=0.007];
	\end{tikzpicture}
	\hspace{-2mm}
}

\foreach \x in {A, ..., Z}{%
	\expandafter\xdef\csname orcid\x\endcsname{\noexpand\href{https://orcid.org/\csname orcidauthor\x\endcsname}{\noexpand\orcidicon}}
}

\newcommand\orcidJonathan{{\href{https://orcid.org/0000-0001-9291-0893}{\orcidicon}}}
\newcommand\orcidEdson{{\href{https://orcid.org/0000-0001-9929-5977}{\orcidicon}}}

\newcommand\orcidAslam{{\href{https://orcid.org/0009-0006-5160-1131}{\orcidicon}}}

\newcommand\orcidFrancisco{{\href{https://orcid.org/0000-0002-9388-8373}{\orcidicon}}}

\usepackage{placeins}

\newcommand{\Del}{\nabla}
\begin{document}

\title{Thin-shell wormholes from magnetic black holes in exponential electrodynamics}
\author{Jonathan A. Rebou\c{c}as\orcidJonathan\!\!}
\email{jalvesreboucas@ifce.edu.br}
\affiliation{Instituto Federal de Educa\c{c}\~ao, Ci\^encia e Tecnologia do Cear\'a (IFCE), Iguatu, CE, Brazil}

\author{Edson Otoniel\orcidEdson\!\!}
\email{edson.otoniel@ufca.edu.br}
\affiliation{Universidade Federal do Cariri (UFCA), Instituto de Forma\c{c}\~ao de Educadores - IFE, R. Oleg\'ario Emidio de Araujo S/N, Brejo Santo - CE, 63.260-000 - Brazil}

\author{Muhammad Israr Aslam\orcidAslam\!\!}
\email{israr.aslam@umt.edu.pk}
\affiliation{Department of Mathematics, School of Science, University of Management and Technology, Lahore-54770, Pakistan}

\author{Francisco S. N. Lobo\orcidFrancisco\!\!} 
\email{fslobo@ciencias.ulisboa.pt}
\affiliation{Instituto de Astrof\'{i}sica e Ci\^{e}ncias do Espa\c{c}o, Faculdade de Ci\^{e}ncias da Universidade de Lisboa, Edifício C8, Campo Grande, P-1749-016 Lisbon, Portugal}
\affiliation{Departamento de F\'{i}sica, Faculdade de Ci\^{e}ncias da Universidade de Lisboa, Edif\'{i}cio C8, Campo Grande, P-1749-016 Lisbon, Portugal}

\begin{abstract}
Transparent magnetic wormholes can share a black hole exterior while supporting matter controls throat response. We construct symmetric thin shells from the magnetic black hole of exponential nonlinear electrodynamics and derive equilibrium stresses and a linear radial stability criterion. The seed's zero-bare-mass magnetic branch ties charge and coupling along the fixed-mass family. In sampled configurations, calibrated linear barotropic shells are unstable, whereas variable phantom and variable Chaplygin laws admit locally stable sectors. Radius dependence strengthens the phantom restoring term; the Chaplygin response cancels the divergent inverse lapse contribution and permits stability near a non-extremal seed horizon. Surface density remains negative, although an inner interval satisfies the tangential null and intrinsic strong energy conditions. At the seed photon sphere, both variable laws share an exact stabilization threshold. The complete radial averaged null energy remains negative even when the proper shell energy tends to zero. Null geodesic imaging uses independent, minimally coupled light and transparent junctions, with celestial sources and rotating thin disks truncated at ISCO. For observer-side disk emission with ISCO truncation and a throat below the seed photon sphere, inclined images coincide with the seed black hole image. A throat above that sphere removes selected returning contributions while preserving much of the bright primary structure. Illuminating the opposite exterior adds inner disk arcs, but a central dark region can survive because the disk does not illuminate every transmitted direction. Image degeneracy depends jointly on retained null trajectories and emitting support, rather than on the presence of a horizon alone.
\end{abstract}

\maketitle

\section{Introduction}\label{sec:introduction}
The first image of M87* released by the Event Horizon Telescope (EHT) established a resolved view of horizon scale emission around a supermassive compact object~\cite{ned1}. Interpreting such an image requires separating the geometry of null trajectories from the distribution and motion of the emitting matter~\cite{ned2}. In particular, a broad bright emission ring is not identical to the narrow photon ring structure associated with near critical trajectories, and a central intensity depression need not coincide with a geometrically defined capture shadow~\cite{GrallaHolzWald2019}. This distinction becomes essential when comparing black holes with horizonless objects that retain part of the same exterior geometry.

From this perspective, traversable wormholes provide a controlled setting in which to examine this issue alongside the material requirements of a throat. The Morris-Thorne construction relates traversability to the geometry and stress supporting the connection~\cite{MorrisThorne1988}. Energy-condition requirements have also been examined in Einstein--Cartan gravity with spin and torsion~\cite{DiGrezia:2017daq} and in generalized-uncertainty-corrected Rastall--Rainbow Casimir wormholes~\cite{Battista:2024gud}. Rotating wormholes in magnetic backgrounds also arise through Harrison transformations in Einstein--Maxwell theory with a conformally coupled scalar~\cite{Cisterna2023Ehlers}. In thin-shell constructions, two exterior regions are joined across a timelike hypersurface~\cite{Visser1989}, and the discontinuity of its extrinsic curvature determines a distributional surface stress through the Israel junction conditions~\cite{Israel1966}. For symmetric spherical junctions in general relativity, the surface density is negative. The response to a radial displacement nevertheless depends on the constitutive properties assigned to the layer. Linearized stability and its dependence on the surface equation of state (EOS) have been developed in both specific and general geometries~\cite{PoissonVisser1995,LoboCrawford2004,LoboCrawford2005,GarciaLoboVisser2012}.

General spherical stability analyses make the density-response dependence explicit~\cite{Eiroa2008}, while regular Hayward seeds provide an early magnetic example with several surface equations of state~\cite{Halilsoy2014}. A particularly useful geometric connection is the relation between circular photon orbits and barotropic thin-shell stability established by Tsukamoto and Kokubu~\cite{TsukamotoKokubu2024}. Earlier work related pure-tension throats to photon spheres~\cite{BarceloVisser2000}, with a geometric extension to photon surfaces in vacuum with cosmological constant $\Lambda$~\cite{Koga2020}. Below we recover the corresponding photon sphere obstruction and identify the additional term supplied by an explicitly radius-dependent surface pressure.

Chaplygin surface fluids and their extensions have long been used to identify locally stable thin-shell configurations~\cite{EiroaSimeone2007,EiroaFigueroa2012}. Explicit dependence of the pressure on the throat radius supplies a further response that is independent of the local density slope. Varela demonstrated this mechanism for Schwarzschild shells, including the distinct near-horizon behavior of variable phantom and variable Chaplygin laws~\cite{Varela2015}. Javed et al. subsequently compared barotropic, variable phantom and variable Chaplygin matter in a nonlinear electrodynamic setting~\cite{Javed2023}. The existence of constitutive stabilization in nonlinear electrodynamics (NLED) is therefore established prior work. The question addressed here is how these responses operate for the exact exponential magnetic lapse, with equilibrium calibration and the inherited mass charge relation treated consistently.

The optical problem has a related dependence on what is retained and illuminated. Wormhole lensing can connect sources and observers in the same or opposite asymptotic regions~\cite{Shaikh2019}. Thin-disk studies have shown that emission on the opposite side can generate additional structures~\cite{Paul2020}, while asymmetric thin-shell constructions exhibit further image branches controlled by their different exterior potentials~\cite{Peng2021}. Charged rotating wormhole images likewise depend on the assumed accretion source~\cite{saleem2024observational}. These results motivate a distinction between a return transmission boundary fixed by the retained geometry and darkness fixed by the absence of supported emission. In the symmetric geometry considered below, illumination of the second exterior changes the observed radiation without changing the null trajectories. Complementary wormhole diagnostics use the Poynting--Robertson effect~\cite{DeFalco:2020afv,DeFalco:2021klh} and reconstruction from modified gravitational potentials~\cite{DeFalco:2021ksd}.

Earlier optical constructions further clarify the role of reflection symmetry. Asymmetric thin-shell spacetimes can have transmission boundaries controlled by the opposite exterior~\cite{Wang2020}, distinct shadow scales associated with their two effective potentials~\cite{Wielgus2020}, and additional disk-image branches in regular Hayward geometries~\cite{Guo2023}. In the identical exterior model studied here, the static symmetric junction preserves the Killing energy, and an accessible transmitted ray cannot be reflected back by a larger potential barrier on the opposite side. These precedents are therefore relevant comparisons rather than interchangeable optical mechanisms.

The present analysis is also related to the multi-fluid and imaging study of a regularized string cloud seed~\cite{Zhong2026Optical} and to the recent construction of magnetic thin shells in Born-Infeld type electrodynamics~\cite{Reboucas2026BornInfeld}. The latter includes different Chaplygin constitutive laws and a thermodynamic analysis. Here we specialize to exponential electrodynamics and compare a calibrated linear barotrope with the radius-dependent phantom and inverse density Chaplygin laws. The source and camera prescriptions are stated explicitly, and observer-side and bilateral illumination are compared at the same geometry. These are complementary model studies, neither the general junction method nor stabilization by variable surface equations of state is claimed as a new construction.

We use the regular magnetic black hole solution of Kruglov~\cite{Kruglov2017}, restricting the seed to its zero bare mass branch. This choice ties the magnetic charge and nonlinear coupling to the asymptotic mass. It is an inherited restriction on the chosen seed family, rather than a general condition imposed by a wormhole junction. The fixed mass scan changes both charge and coupling, whereas each individual radial perturbation keeps the bulk theory and its parameters fixed. The optical probe is a separate, minimally coupled test field. Its trajectories are metric null geodesics and are not identified with the characteristic rays of perturbations of the supporting nonlinear electromagnetic field~\cite{Novello2000,MurkSoranidis2024}.

Within these assumptions we derive the equilibrium stresses, the general radial restoring curvature, and its specialization to the three surface laws. We then compare celestial source maps and images of circularly moving thin accretion disks truncated at the innermost stable circular orbit (ISCO). The scope is local linear radial stability and stationary geometric optics appearance; no microscopic shell model, full coupled perturbative stability, or self-consistent accretion solution is assumed. The manuscript is organized as follows: Section~\ref {sec:exponential} specifies the seed and admissible exterior. Sections~\ref{sec:magnetic_throat} and~\ref{sec:surface_stresses} give the construction and energy inequalities of thin-shell wormholes, respectively. Section~\ref {sec:radial_mode} investigates junction conservation, the dynamics of reversible thermodynamics, and the corresponding thermal response. Section~\ref{sec:constitutive} analyzes the linear barotropic, variable phantom, and variable Chaplygin surface laws and their linear radial stability. The equilibrium of the photon sphere, explicit radius stabilization, and the radial support of the imaged throat are also investigated in the same section. Section~\ref{sec:optical_appearance} presents the optical prescription and the resulting thin-disk images. Finally, Sec.~\ref {sec:conclusion} summarizes our results and discusses possible future extensions.

\section{Exponential magnetic geometry and its exterior domain}\label{sec:exponential}

We use the magnetic solution of exponential nonlinear electrodynamics obtained by Kruglov~\cite{Kruglov2017}, with no independent bare mass. In geometrized units, $G=c=1$, with $G$ the gravitational constant and $c$ the speed of light, its line element can be written directly in reduced coordinates as:
\begin{equation}
 \begin{split}
 ds^2=\ell^2\biggl[&-f(\xi)\,dT^2+\frac{d\xi^2}{f(\xi)}\\
 &+\xi^2\left(d\theta^2+\sin^2\theta\,d\phi^2\right)\biggr].
 \end{split}
 \label{eq:magnetic_metric}
\end{equation}
Here $\ell>0$ is a length scale, $\xi=r/\ell$ is the dimensionless areal coordinate, and $T=t/\ell$ is the dimensionless static time. The physical coordinates $r$ and $t$ have dimensions of length, $\theta$ and $\phi$ are the polar and azimuthal angles, respectively. The signature is $(-,+,+,+)$. The coordinate is areal, so this geometry has spherical area $4\pi\ell^2\xi^2$, a fact that directly controls the angular junction curvature and the dilution of the shell density.

The lapse function defines the complete magnetic mass profile as given below
\begin{equation}
 f(\xi)=1-\frac{\Gamma\!\left(\frac14,\xi^{-4}\right)}{\lambda\xi}.
 \label{eq:exponential_lapse}
\end{equation}
The parameter $\lambda>0$ is dimensionless and measures the shape of the geometry after the radial reduction. The upper incomplete gamma function is defined by $\Gamma(s,z)=\int_z^\infty v^{s-1}e^{-v}\,dv$, with auxiliary arguments $s$ and $z$ and integration variable $v$; $\Gamma(s)$ denotes its complete limit at $z=0$. This convention is important because the lower incomplete function would describe a different radial profile. The reduction is related to the magnetic charge and nonlinear coupling by
\begin{equation}
 \ell=\left(\frac{\beta_gQ^2}{2}\right)^{1/4},
 \qquad \lambda=\frac{2\sqrt{2\beta_g}}{|Q|}.
 \label{eq:magnetic_reduction}
\end{equation}
The geometrized charge $Q$ and positive coupling $\beta_g$ carry length and length-squared dimensions, respectively. The metric is unchanged under $Q\mapsto-Q$. The original dimensionless radius and parameter in Ref.~\cite{Kruglov2017} are thus represented here by $\xi$ and $\lambda$, respectively. Before setting $G=1$, their relation to the natural unit quantities of that work is $Q=\sqrt G\,q$ and $\beta_g=\beta/G$, where $q$ and $\beta$ are its magnetic charge and nonlinear coupling.

We normalize the electromagnetic matter Lagrangian as $\mathcal L_{\rm m}=-\mathcal F\allowbreak\exp(-\beta_g\mathcal F)/(4\pi)$, together with the Einstein-Hilbert density $R/(16\pi)$. Here $\mathcal F=F_{\mu\nu}F^{\mu\nu}/4$, $F_{\mu\nu}$ is the electromagnetic field tensor, $R$ is the Ricci scalar, and Greek indices label the four bulk coordinates. The factor $1/(4\pi)$ fixes the gravitational normalization so that the mass profile in Eq.~\eqref{eq:exponential_lapse} has the usual Maxwell charge term. For the purely magnetic branch the energy density is $\rho=\mathcal F e^{-\beta_g\mathcal F}/(4\pi)$, and the mass balance is $dm/dr=4\pi r^2\rho$. The function $m(r)$ is the enclosed geometrical mass, not an additional free parameter. This normalization reproduces the stated metric and Maxwell asymptotic term; all subsequent junction and perturbation equations use the same gravitational convention.

In the reduced coordinate, the magnetic invariant satisfies
\begin{equation}
 \mathcal F(\xi)=\frac{Q^2}{2\ell^4\xi^4},
 \qquad \beta_g\mathcal F(\xi)=\xi^{-4}.
 \label{eq:magnetic_invariant}
\end{equation}
The invariant has dimensions of inverse length squared. Its exponential suppression makes the integrated magnetic energy finite, although the magnetic field itself grows toward the origin. Integrating the mass balance from the center, without a mass integration constant, gives
\begin{equation}
 \mu\equiv\frac{M}{\ell}=\frac{\Gamma(1/4)}{2\lambda}.
 \label{eq:magnetic_mass}
\end{equation}
The total geometrical mass $M=m(\infty)$ is the mass measured at either asymptotic end, and $\mu$ is its reduced value. Consequently, the mass, charge, and nonlinear scale cannot all be varied independently within this branch. Equations~\eqref{eq:magnetic_reduction} and \eqref{eq:magnetic_mass} imply $Q^2/\ell^2=4/\lambda$ and $\beta_g/\ell^2=\lambda/2$. At fixed $M$, they instead give $\ell/M=2\lambda/\Gamma(1/4)$, $Q^2/M^2=16\lambda/\Gamma(1/4)^2$, and $\beta_g/M^2=2\lambda^3/\Gamma(1/4)^2$. These dependencies distinguish a change of geometric shape from a change of overall length scale. Because $\beta_g$ is a coupling of the electromagnetic Lagrangian, changing $\lambda$ at fixed $M$ compares different coupling choices as well as different charges. At fixed $M$ and fixed $\beta_g$, the zero-bare-mass branch instead fixes $\lambda$ and $|Q|$. None of the parameter-family figures represents the evolution of one shell or an isolated charge variation within a fixed theory.

The mass relation in Eq.~\eqref{eq:magnetic_mass} is inherited from the regular seed by integrating its magnetic energy from the center. The wormhole removes that center and the interior region. Its exterior metric is deliberately chosen from this seed branch; the junction conditions alone would not enforce the same relation for a more general magnetic exterior with an independent mass integration constant. The Arnowitt--Deser--Misner (ADM) mass of the joined spacetime includes the gravitational effect of both the retained bulk and the shell. It should not be equated to an integral of magnetic energy over a regular center that is absent from the wormhole.

At large radius, the lapse function approaches the Reissner-Nordstr\"om form, with the leading nonlinear correction at inverse sixth order,
\begin{equation}
 f(\xi)=1-\frac{2\mu}{\xi}+\frac{4}{\lambda\xi^2}
 -\frac{4}{5\lambda\xi^6}+\mathcal O(\xi^{-10}).
 \label{eq:magnetic_asymptotic}
\end{equation}
This expansion establishes asymptotic flatness and the magnetic-charge normalization. It also separates the ordinary charge contribution from the genuinely exponential correction. The formal central limit has $f\to1$, with exponentially small departures, so the geometry has a locally Minkowski core in the zero bare mass branch. This geometric regularity does not establish that electromagnetic perturbations remain admissible in the strong field core. That region is removed in the wormhole construction. A regular charged center is excluded for globally causal, asymptotically flat spherical Einstein--NLED solutions with a weak-field expansion~\cite{RussoTownsend2026CausalBH}; the excised thin-shell geometry is not a counterexample.

Regular magnetic solutions of nonlinear electrodynamics have a broader theoretical context~\cite{Bronnikov2001}. Regularity of a background metric and local admissibility of its electromagnetic constitutive derivatives are distinct from stability of coupled perturbations. The latter has been studied for self-gravitating nonlinear electrodynamics~\cite{MorenoSarbach2003} and, specifically, for magnetic black holes~\cite{Nomura2020}. Central Laplacian instabilities of nonsingular nonlinear electrodynamics black holes provide a further reason not to infer physical stability from bounded background curvature~\cite{DeFeliceTsujikawa2025}. The central region is excised in the present wormhole, so that result does not by itself demonstrate an instability of the retained domain. Conversely, black hole perturbation criteria cannot certify stability of the joined geometry without the perturbative boundary conditions at the shell. The positive horizons are determined by the transcendental equation
\begin{equation}
 \Gamma\!\left(\frac14,\xi_h^{-4}\right)=\lambda\xi_h.
 \label{eq:magnetic_horizon}
\end{equation}
The symbol $\xi_h$ denotes a positive root. There are two roots, an inner Cauchy horizon $\xi_-$ and an outer event horizon $\xi_+$, for $0<\lambda<\lambda_e$. At the critical value they merge at $\xi_e$. The critical pair is characterized by
\begin{equation}
 \lambda_e=\frac{\Gamma\!\left(\frac14,\xi_e^{-4}\right)}{\xi_e},
 \qquad \xi_e\Gamma\!\left(\frac14,\xi_e^{-4}\right)=4e^{-\xi_e^{-4}}.
 \label{eq:magnetic_extremal}
\end{equation}
Numerically, $\xi_e\simeq2.1453752260$ and $\lambda_e\simeq0.8289964051$. For $\lambda>\lambda_e$ the lapse has no horizon. We retain the non-extremal black hole sector and its exterior $\xi>\xi_+(\lambda)$, where $f(\xi)>0$ and the lapse increases outward. The rounded value $0.83$ lies above the critical value and belongs to the horizonless sector. The endpoint $\lambda=0$ is singular in Eq.~\eqref{eq:exponential_lapse}; the Schwarzschild limit is obtained at fixed physical $M$ by letting $Q$ and $\beta_g$ vanish together, rather than by substituting zero into the reduced lapse. The appropriate radial ratio in that limit is $\xi/\mu=r/M$, since $\ell/M$ also vanishes. At fixed nonzero $Q$, the Maxwell limit of the electromagnetic Lagrangian does not preserve a finite mass generated solely by the field. On a purely magnetic background, the complete characteristic conditions~\cite{Schellstede2016Causality,RussoTownsend2024Born} are $\mathcal L_{{\rm m},\mathcal F}<0$, $\mathcal L_{{\rm m},\mathcal F\mathcal F}\geq0$ and $\mathcal L_{{\rm m},\mathcal F}+2\mathcal F\mathcal L_{{\rm m},\mathcal F\mathcal F}<0$, where commas denote differentiation with respect to $\mathcal F$. For the electromagnetic admissibility conditions adopted in Ref.~\cite{Kruglov2017}, the magnetic exterior obeys
\begin{equation}
 \begin{gathered}
 \beta_g\mathcal F<y_*\equiv\frac{5-\sqrt{17}}4,\\
 \xi>\xi_*\equiv y_*^{-1/4}\simeq1.4614305338.
 \end{gathered}
 \label{eq:magnetic_field_domain}
\end{equation}
The constants $y_*$ and $\xi_*$ mark the common boundary of the derivative inequalities for the exponential electromagnetic Lagrangian. Strict inequalities leave a neighborhood available for perturbations. Since $\xi_+\geq\xi_e>\xi_*$, a throat strictly outside the event horizon automatically respects this local magnetic restriction. The removed core may lie outside that restriction even though its metric is regular. No numerical bound on the dimensionless $\lambda$ follows from a laboratory field bound without specifying a physical mass and converting the electromagnetic units. These conditions hold throughout every retained exterior and for sufficiently small radial throat perturbations. They do not establish global causality of the exponential theory, whose convexity and strong-field causality fail on other allowed electromagnetic backgrounds~\cite{Schellstede2016Causality,RussoTownsend2024Born}.

\begin{figure}[tbp]
    \begin{minipage}{\columnwidth}
    \centering
    \includegraphics[width=\linewidth]{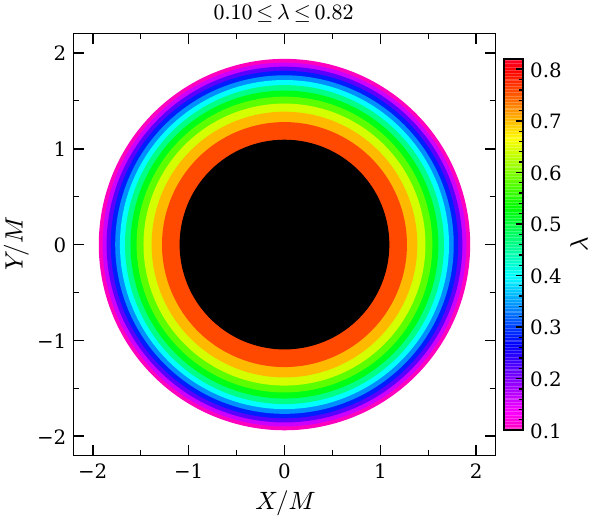}
    \caption{Outer event horizon family of the exponential magnetic black hole seed at fixed mass $M$, for $0.10\leq\lambda\leq0.82$. Thirteen equally spaced parameter values are represented by nested disks; the color bar denotes the dimensionless parameter $\lambda$. The auxiliary equatorial coordinates are $X=r\cos\phi$ and $Y=r\sin\phi$, and the radius of each circle is $r_+(\lambda)/M$. This is a parameter-family diagram. The black disk covers the smallest selected outer horizon radius and is a graphical convention, rather than an additional horizon or an optical shadow. These seed horizons are excised in the thin-shell construction.}
    \label{fig:exponential-outer-horizon-family}
    \end{minipage}
\end{figure}

Figure~\ref{fig:exponential-outer-horizon-family} shows a monotonic decrease of the outer event-horizon radius over the displayed parameter interval. The endpoint values are $r_+/M\simeq1.93717$ at $\lambda=0.10$ and $r_+/M\simeq1.08839$ at $\lambda=0.82$. The mass normalization is essential to this comparison: Eq.~\eqref{eq:magnetic_mass} fixes the conversion from the reduced horizon coordinate to a physical radius, while the dependencies following that equation require the charge and nonlinear coupling to change together as $\lambda$ varies. At a fixed radius in units of $M$, increasing $\lambda$ increases the lower argument of the upper incomplete gamma function in Eq.~\eqref{eq:exponential_lapse}. Its decrease reduces the fraction of the total electromagnetic mass enclosed within that radius, raising the lapse there. Since the outer branch has positive radial lapse slope, its zero shifts inward. The contraction therefore reflects a change in the magnetic mass distribution at fixed asymptotic mass.

The wider inner color bands correspond to larger changes in the outer radius between successive, equally spaced parameter values near the upper end of the interval. The last configuration remains below the critical value defined by Eq.~\eqref{eq:magnetic_extremal}, with distinct inner and outer horizons; the finite central radius does not mark a disappearance of the event horizon. For the symmetric construction, the decreasing outer radius lowers the geometric bound on the physical throat radius in units of $M$, so smaller throat radii become accessible along this family. The throat must still remain strictly outside the relevant horizon. Its surface stresses and restoring response require the junction and constitutive analysis; their signs cannot be inferred from the horizon radius alone.

\section{Identical exteriors and the magnetic throat}\label{sec:magnetic_throat}

The wormhole joins two copies of the exterior in Eq.~\eqref{eq:magnetic_metric}, with the same $\ell$ and $\lambda$, each restricted to $\xi\geq a(u)>\xi_+$. The dimensionless radius $a$ determines the physical throat radius $\ell a$, and $u=\tau/\ell$ is reduced proper time, with $\tau$ the shell proper time. The boundaries are identified at equal $u,\theta,\phi$, forming the timelike throat $\Sigma$. This excises both horizons and the central region from the retained spacetime. The radius increases away from the identified surface toward either infinity, so its area is a minimum on a spatial slice of the joined geometry. In the intrinsic coordinates $y^i=(u,\theta,\phi)$ the two embeddings are
\begin{equation}
 X_\pm^\mu(y^i)=\bigl(T_\pm(u),a(u),\theta,\phi\bigr).
 \label{eq:magnetic_embedding}
\end{equation}
The signs label the two copies, $X_\pm^\mu$ maps the intrinsic chart into each bulk chart, and Latin indices refer to the three shell coordinates. Equality of the angular and temporal pullbacks implements reflection symmetry. In particular, the first fundamental form is obtained from
\begin{equation}
 h_{ij}=g_{\mu\nu}\,
 \frac{\partial X_\pm^\mu}{\partial y^i}
 \frac{\partial X_\pm^\nu}{\partial y^j}.
 \label{eq:magnetic_pullback}
\end{equation}
Here $h_{ij}$ is the induced metric and $g_{\mu\nu}$ is the bulk metric of Eq.~\eqref{eq:magnetic_metric}. Proper-time parametrization imposes $f(a)\dot T_\pm^2-\dot a^2/f(a)=1$, where dots denote differentiation with respect to $u$; the future directed solution is $\dot T_\pm=\sqrt{f(a)+\dot a^2}/f(a)$. Substitution into the pullback gives
\begin{equation}
 ds_\Sigma^2=\ell^2\left[-du^2+a(u)^2
 \left(d\theta^2+\sin^2\theta\,d\phi^2\right)\right].
 \label{eq:magnetic_induced}
\end{equation}
Thus the shell area is $4\pi\ell^2a^2$, and the time direction remains timelike as long as the motion stays in the retained exterior. The two copies share the same induced metric even when the boundary moves. The trajectory has a local level set representation $\Phi_\pm(T_\pm,\xi)=\xi-a[u(T_\pm)]$. The oriented unit normals are
\begin{equation}
 \begin{aligned}
 n_\mu^{(\pm)}&=
 \pm\left|g^{\alpha\beta}\partial_\alpha\Phi_\pm
 \partial_\beta\Phi_\pm\right|^{-1/2}\partial_\mu\Phi_\pm
 \\&=\pm\ell\left(-\dot a,\frac{\sqrt{f(a)+\dot a^2}}{f(a)},0,0\right).
 \end{aligned}
 \label{eq:magnetic_normal}
\end{equation}
The co-vector $n_\mu^{(\pm)}$ has unit spacelike norm and is orthogonal to the shell trajectory. Its relative sign gives opposite radial orientations in the two exterior charts. The dependence on $u(T_\pm)$ expresses the moving boundary in bulk coordinates and enters its normal through differentiation of the level set. Our curvature convention is
\begin{equation}
 K_{ij}^{(\pm)}=
 \frac{\partial X_\pm^\mu}{\partial y^i}
 \frac{\partial X_\pm^\nu}{\partial y^j}
 \Del_\mu n_\nu^{(\pm)}.
 \label{eq:magnetic_extrinsic_definition}
\end{equation}
The tensor $K_{ij}^{(\pm)}$ is the second fundamental form, and $\Del$ is the covariant derivative appropriate to the indicated metric. In an orthonormal frame of the shell, marked by hatted indices, its independent covariant components are
\begin{equation}
 \begin{aligned}
 K_{\hat u\hat u}^{(\pm)}&=
 \mp\frac{f'(a)+2\ddot a}{2\ell\sqrt{f(a)+\dot a^2}},
 \\[1ex]
 K_{\hat\theta\hat\theta}^{(\pm)}&=
 K_{\hat\phi\hat\phi}^{(\pm)}=
 \pm\frac{\sqrt{f(a)+\dot a^2}}{\ell a}.
 \end{aligned}
 \label{eq:magnetic_extrinsic_components}
\end{equation}
Primes on $f$ denote differentiation with respect to its dimensionless argument. Both components have dimensions of inverse length. The angular jump arises because the areal radius grows away from the throat on both sides; the temporal component also contains the radial acceleration. Lowering a timelike orthonormal index produces the relative temporal sign displayed here.

The magnetic field across $\Sigma$ is described, with the same angular identification, by the continuous two-form $\boldsymbol F=Q\sin\theta\,d\theta\wedge d\phi$, where $\boldsymbol F$ is the differential-form representation of $F_{\mu\nu}$. Its pullback is identical on the two sides and its contraction with the shell normal vanishes, even during radial motion. The magnetic Bianchi identity and the normal electromagnetic junction equation therefore introduce neither a magnetic charge sheet nor an electric surface current. The flux measured with outward orientation at one infinity is opposite to that measured at the other. The geometry remains reflection symmetric because the magnetic invariant depends on $Q^2$; equal oriented charges at both infinities would require a different electromagnetic junction.

\section{Surface stresses and energy inequalities}\label{sec:surface_stresses}

The matter concentrated on the throat is an isotropic surface fluid. We write its covariant orthonormal tensor as
\begin{equation}
 S_{\hat i\hat j}=\frac1\ell\operatorname{diag}(\sigma,p,p).
 \label{eq:magnetic_surface_tensor}
\end{equation}
The surface tensor $S_{\hat i\hat j}$ is physical, while $\sigma$ and $p$ are the reduced density and tangential pressure; their physical values are $\sigma/\ell$ and $p/\ell$, respectively. Thus $\sigma$ and $p$ are dimensionless. The conservation law and constitutive relations use the same reduced stresses. The orthonormal metric is $\eta_{\hat i\hat j}=\operatorname{diag}(-1,1,1)$; in mixed components the temporal entry of $S$ is consequently negative. The Israel-Lanczos condition~\revMagnetic{\cite{Israel1966,PoissonVisser1995,GarciaLoboVisser2012}} in the same basis is
\begin{equation}
 S_{\hat i\hat j}=-\frac1{8\pi}
 \left([K_{\hat i\hat j}]-\eta_{\hat i\hat j}[K]\right).
 \label{eq:magnetic_lanczos}
\end{equation}
Brackets denote the difference of the plus and minus limits, and $[K]=\eta^{\hat i\hat j}[K_{\hat i\hat j}]$ is their trace in this basis. Using Eq.~\eqref{eq:magnetic_extrinsic_components}, the angular jump fixes
\begin{equation}
 \sigma=-\frac{\sqrt{f(a)+\dot a^2}}{2\pi a}.
 \label{eq:magnetic_dynamic_density}
\end{equation}
The negative sign follows from the orientation of the joined exteriors. It is not a constitutive assumption, and changing an equation of state cannot reverse it within this symmetric construction. The temporal jump gives
\begin{equation}
 p=\frac1{8\pi}\left[
 \frac{f'(a)+2\ddot a}{\sqrt{f(a)+\dot a^2}}
 +\frac{2\sqrt{f(a)+\dot a^2}}{a}\right].
 \label{eq:magnetic_dynamic_pressure}
\end{equation}
The pressure couples the acceleration of the throat to both the lapse gradient and its angular curvature. Radial changes in the surface fluid therefore affect the motion through this junction equation, rather than through an independently prescribed mechanical force. At a static throat, $a=a_0>\xi_+$, the reduced stresses become
\begin{equation}
 \sigma_0=-\frac{\sqrt{f(a_0)}}{2\pi a_0},
 \qquad
 p_0=\frac{a_0f'(a_0)+2f(a_0)}{8\pi a_0\sqrt{f(a_0)}}.
 \label{eq:magnetic_static_stresses}
\end{equation}
The subscript zero indicates evaluation at equilibrium. Since the exterior has $f>0$ and $f'>0$, one has $\sigma_0<0$ and $p_0>0$. The weak and dominant energy conditions of the layer fail because each requires nonnegative density. The bulk magnetic matter and the surface fluid are distinct sources: the magnetic exterior may satisfy its local energy inequalities while the topological junction still requires a negative surface density. For tangent null directions, the null energy condition (NEC) is governed by
\begin{equation}
 \sigma_0+p_0=
 \frac{a_0f'(a_0)-2f(a_0)}{8\pi a_0\sqrt{f(a_0)}}.
 \label{eq:magnetic_tangent_nec}
\end{equation}
Thus a positive lapse slope is necessary but is not sufficient for the tangential NEC: it must overcome the angular dilution term. The separate combination
\begin{equation}
 \sigma_0+2p_0=\frac{f'(a_0)}{4\pi\sqrt{f(a_0)}}
 \label{eq:magnetic_trace_combination}
\end{equation}
is positive throughout this exterior, but its positivity does not imply the intrinsic strong energy condition (SEC) of a layer with two spatial directions. In three-dimensional world-volume geometry, the SEC is tested with $S_{ij}-h_{ij}\operatorname{tr}S$, where $\operatorname{tr}S=h^{kl}S_{kl}$ is the surface trace; for an isotropic fluid it requires $p_0\geq0$ together with $\sigma_0+p_0\geq0$. The first requirement holds here, so the intrinsic SEC has the same radial restriction as the tangential NEC. In this regard, Eq.~\eqref{eq:magnetic_trace_combination} remains a useful stress diagnostic, not an alternative definition of that condition.

A null vector crossing the shell in a radial direction has no angular component. Its contraction with the distributional surface stress is proportional to the negative density. Hence fulfillment of the tangential NEC on part of the throat family does not remove the violation of the four-dimensional null condition at the shell. The construction remains supported by exotic surface matter.

Close to a simple outer horizon, $f(a_0)$ is proportional to $a_0-\xi_+$ and $f'(\xi_+)>0$. The density then vanishes from below as the square root of the separation, whereas the pressure diverges as its inverse square root. The tangential NEC is positive sufficiently close to that horizon, but the shell is not matter-free: its pressure becomes singular. At large $a_0$, Eq.~\eqref{eq:magnetic_asymptotic} gives $\sigma_0\sim-1/(2\pi a_0)$ and $p_0\sim1/(4\pi a_0)$, so the tangential NEC eventually fails. Its zeros are determined by the numerator of Eq.~\eqref{eq:magnetic_tangent_nec}; the pressure and density alone do not establish their number. A degenerate horizon requires a separate limiting expansion and is not a timelike equilibrium point.

\begin{figure}[tbp]
    \begin{minipage}{\columnwidth}
    \centering
    \includegraphics[width=\linewidth]{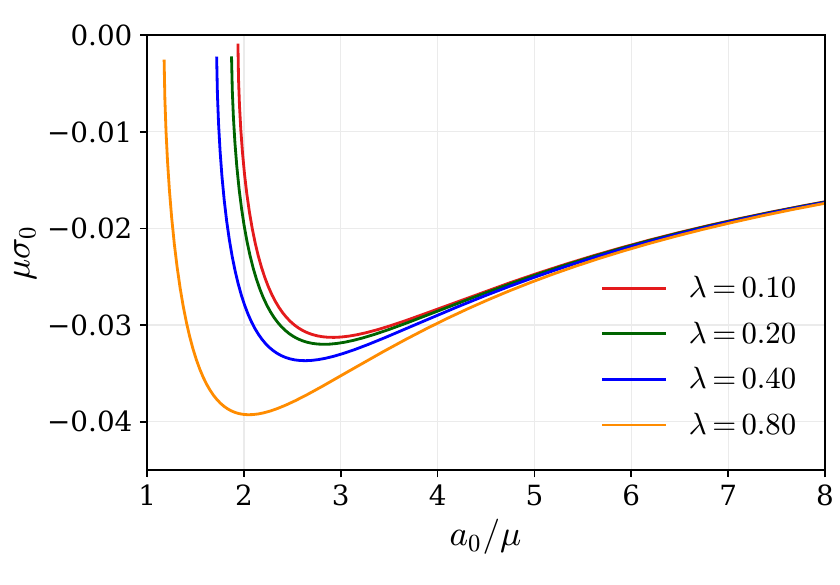}
    \caption{Mass-normalized static surface density $\mu\sigma_0$ as a function of the mass-normalized equilibrium throat radius $a_0/\mu$ for the symmetric exponential magnetic thin-shell wormhole. The four curves have $\lambda=0.10,0.20,0.40,0.80$, with fixed asymptotic mass $M$ and $\mu$ given by Eq.~\eqref{eq:magnetic_mass}. The surface density is evaluated from Eq.~\eqref{eq:magnetic_static_stresses}; $\mu\sigma_0$ is the physical surface density multiplied by $M$, while $a_0/\mu$ is the physical throat radius divided by $M$. Each curve includes only the static exterior $a_0>\xi_+(\lambda)$ and obeys the local magnetic restriction in Eq.~\eqref{eq:magnetic_field_domain}. The numerical grid has 3600 points over $1\leq a_0/\mu\leq8$; points at or below the outer horizon are excluded.}
    \label{fig:exponential-shell-density}
    \end{minipage}
\end{figure}

The density profiles in Fig.~\ref{fig:exponential-shell-density} have a single negative minimum within the displayed exterior interval. Increasing $\lambda$ moves that minimum toward smaller mass-normalized throat radii and increases its magnitude. Between $\lambda=0.10$ and $0.80$, its location changes from $a_0/\mu\simeq2.91653$ to $2.05151$, while its value changes from $\mu\sigma_0\simeq-0.03128$ to $-0.03928$. At any common radius in the exterior of all four seeds, the curves remain ordered, with the larger parameter giving the more negative density. This dependence follows from Eq.~\eqref{eq:magnetic_static_stresses}: the increase of the lapse at a fixed mass-normalized radius discussed for Fig.~\ref{fig:exponential-outer-horizon-family} increases the square-root factor in the density, while the physical areal dilution factor remains unchanged. The displaced left endpoints reflect the different outer-horizon radii; the first finite samples lie strictly outside those boundaries.

The nonmonotonic radial dependence results from competition between the outward increase of the lapse and dilution by the areal radius. Differentiating the density in Eq.~\eqref{eq:magnetic_static_stresses} shows that its stationary points coincide with zeros of the numerator in Eq.~\eqref{eq:magnetic_tangent_nec}. Thus the density minimum of each displayed branch also marks saturation of the tangential NEC, separating its positive near-horizon combination from its negative outer portion. This geometric correspondence does not remove the negative density itself. Beyond the minima, the profiles approach the common leading inverse-radius behavior established by Eqs.~\eqref{eq:magnetic_asymptotic} and \eqref{eq:magnetic_static_stresses}, and their parameter-dependent separation becomes smaller. These profiles describe the equilibrium stresses shared by the three constitutive laws when calibrated at the same throat; a minimum of the density is not a criterion for radial stability, which depends on the pressure response to a displacement.

\begin{figure*}[t]
    \begin{minipage}{\textwidth}
    \centering
    \includegraphics[width=\linewidth]{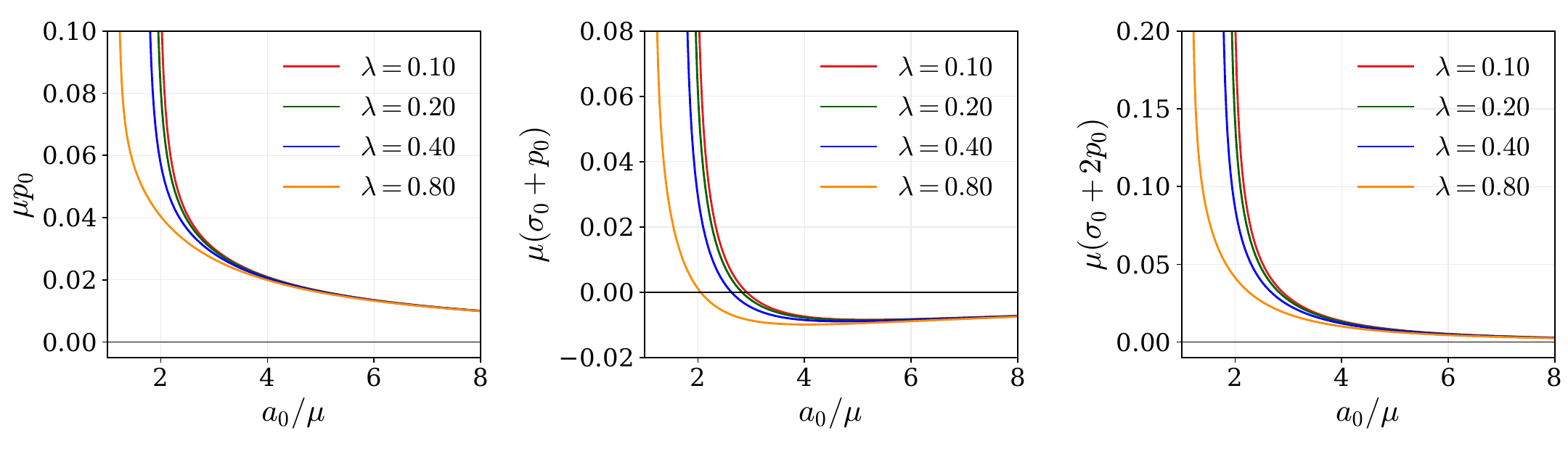}
    \caption{Mass-normalized static pressure $\mu p_0$, tangential NEC combination $\mu(\sigma_0+p_0)$, and stress combination $\mu(\sigma_0+2p_0)$ as functions of $a_0/\mu$, from left to right. The third combination is a stress diagnostic rather than the intrinsic SEC. The four curves have $\lambda=0.10,0.20,0.40,0.80$ at fixed asymptotic mass $M$, with $\mu$ given by Eq.~\eqref{eq:magnetic_mass}. Each ordinate equals the corresponding physical surface stress multiplied by $M$; $a_0/\mu$ is the physical throat radius divided by $M$. The profiles follow Eqs.~\eqref{eq:magnetic_static_stresses}, \eqref{eq:magnetic_tangent_nec} and \eqref{eq:magnetic_trace_combination}. Only the static exterior $a_0>\xi_+(\lambda)$ satisfying Eq.~\eqref{eq:magnetic_field_domain} is shown. The horizontal black line denotes zero. Positive divergences at the nonextremal outer horizons extend above the displayed vertical windows.}
    \label{fig:exponential-shell-pressure-energy}
    \end{minipage}
\end{figure*}

The stress profiles in Fig.~\ref{fig:exponential-shell-pressure-energy} distinguish the positive pressure required at equilibrium from the sign of its sum with the negative density. The pressure decreases throughout the displayed exterior interval, whereas the third combination remains positive and decays more rapidly at larger radii. Their positivity follows from the positive lapse and lapse gradient in Eqs.~\eqref{eq:magnetic_static_stresses} and \eqref{eq:magnetic_trace_combination}. Close to each non-extremal outer horizon, the lapse-gradient contribution dominates all three profiles. Their rise beyond the plotted vertical windows represents the singular pressure limit described above, rather than a finite maximum or an admissible equilibrium on the horizon.

The middle panel resolves the finite radial interval in which the pressure compensates the negative density in tangent null contractions. For increasing values of $\lambda$ in the displayed sequence, the zeros occur at $a_0/\mu\simeq2.91653,\ 2.82783,\ 2.62972,$ and $2.05151$, respectively. These are the density-minimum radii identified in Fig.~\ref{fig:exponential-shell-density}. The numerator in Eq.~\eqref{eq:magnetic_tangent_nec} is positive between the relevant outer horizon and this zero, and negative beyond it for the four configurations considered. Both boundaries of the positive interval move inward as $\lambda$ increases, while its width in units of $M$ decreases from approximately $0.97937$ to $0.87795$ between the first and last curves. Because the pressure is positive, the intrinsic SEC has the same interval of satisfaction as the tangential NEC. The persistent positivity of the third panel therefore does not extend that interval. The negative density in Fig.~\ref{fig:exponential-shell-density} still excludes the weak and dominant conditions throughout these branches.

Beyond the zero, the tangential combination develops a negative minimum before returning toward zero from below. Across the sampled family, increasing $\lambda$ shifts this minimum to smaller radii and makes it more negative. At large radius, the asymptotic lapse in Eq.~\eqref{eq:magnetic_asymptotic} accounts for the decreasing separation of the curves: the density and pressure have parameter-independent leading inverse-radius terms in mass units. In the third combination these terms cancel, leaving an inverse-square leading decay, as follows from Eq.~\eqref{eq:magnetic_trace_combination}. Its faster dilution explains why that panel approaches zero while the middle panel retains a visible negative tail. The common asymptotic behavior does not imply identical stresses at finite throat radius.

\subsection{Proper shell energy and radial averaged null energy}
\label{subsec:radial_anec}
The amount of exotic matter depends on the measure being considered~\cite{VisserKarDadhich2003}. To distinguish the vanishing near-horizon surface density from a null geodesic average, we now use physical quantities. Let $r_0=\ell a_0$ be the static physical throat radius, $f_0=f(a_0)>0$, and $\sigma_{\rm phys,0}=\sigma_0/\ell$. The integral of the density over the proper shell area is

\begin{equation}
 E_{\rm shell}=4\pi r_0^2\sigma_{\rm phys,0}
 =-2r_0\sqrt{f_0}.
 \label{eq:shell_proper_energy}
\end{equation}

This is the proper energy of the layer, not its contribution in isolation to the ADM mass. At fixed seed parameters it tends to zero from below as $a_0\to\xi_+^+$, even though the physical pressure diverges at a simple horizon. Consider a complete future-directed radial null geodesic from one asymptotic infinity to the other, with affine parameter $s$ and tangent $k^\mu=dx^\mu/ds$. Define its conserved Killing energy by $E=-k_\mu(\partial_t)^\mu>0$, using the physical asymptotic time $t=\ell T$. The common clock normalization of the static symmetric junction gives the same $E$ on the two sides. In a physical Gaussian normal coordinate $z$ through the throat, with the shell at $z=0$, let $T_{\mu\nu}=T^{\rm bulk}_{\mu\nu}+T^{\rm shell}_{\mu\nu}$ denote the total stress tensor. The distributional shell stress and the local null components satisfy

\begin{equation}
 \begin{gathered}
 T^{\rm shell}_{\mu\nu}k^\mu k^\nu
 =\sigma_{\rm phys,0}\frac{E^2}{f_0}\,\delta(z),\\
 k^{\hat\tau}=\frac{E}{\sqrt{f_0}},\qquad
 \left|\frac{dz}{ds}\right|_{\Sigma}=\frac{E}{\sqrt{f_0}}.
 \end{gathered}
 \label{eq:anec_shell_distribution}
\end{equation}

Here $\delta(z)$ is the Dirac distribution. The angular pressure does not enter this contraction because the geodesic is radial. In each smooth bulk region the radial magnetic pressure is $p_r=-\rho$, so the contribution is zero: $T^{\rm bulk}_{\mu\nu}k^\mu k^\nu=(\rho+p_r)(k^{\hat t})^2=0$ for the magnetic source. The entire complete geodesic average therefore comes from the single shell crossing, giving

\begin{equation}
 \begin{split}
 \mathcal A_{\rm rad}
 &\equiv\int_{-\infty}^{+\infty}
 T_{\mu\nu}k^\mu k^\nu\,ds\\
 &=\sigma_{\rm phys,0}\frac{E}{\sqrt{f_0}}
 =-\frac{E}{2\pi r_0}<0.
 \end{split}
 \label{eq:radial_anec_exact}
\end{equation}

For definiteness, one may set $E=1$ at infinity. A positive affine rescaling changes the numerical magnitude of $\mathcal A_{\rm rad}$ but not its sign. With that normalization, the limit at a fixed non-extremal seed is $\mathcal A_{\rm rad}\to-1/(2\pi\ell\xi_+)$, which remains finite and negative. Thus $E_{\rm shell}\to0$ does not mean that the radial averaged null energy condition (ANEC) violation disappears. The limiting surface itself is not an admissible timelike shell; these statements concern exterior limits of the static family. No quantum source for the layer is assumed in this calculation. Were its negative energy attributed to quantum fields, additional quantum inequality restrictions would have to be examined~\cite{FewsterRoman2005}. Those restrictions are not consequences already derived for the phenomenological classical surface fluids used here.

\section{Conservation, radial dynamics, and thermodynamics}\label{sec:radial_mode}

\subsection{Junction conservation and linear radial response}
\label{subsec:magnetic_radial_response}

The contracted junction identity \revMagnetic{\cite{LoboCrawford2005,GarciaLoboVisser2012}} relates changes in the layer to the normal flux from the bulk,
\begin{equation}
 \Del_i S^i{}_j=-[T_{\mu\nu}n^\mu e^\nu{}_j].
 \label{eq:magnetic_flux_identity}
\end{equation}
Here $T_{\mu\nu}$ is evaluated on the smooth bulk sides, where it equals the electromagnetic tensor $T^{\rm bulk}_{\mu\nu}$, and $e^\nu{}_j=\partial X^\nu/\partial y^j$ is a tangent vector to the shell; the left-hand derivative is the intrinsic one. A radial magnetic configuration has radial pressure $p_r=-\rho$. A radial boost therefore produces no energy flux along the shell trajectory, and the contraction on the right vanishes on each side. This is a property of the bulk stress, not merely a consequence of identical metric coefficients. Since the true area is proportional to $a^2$, the reduced density obeys
\begin{equation}
 \frac{d\sigma}{da}=-\frac2a\,[\sigma+p(\sigma,a)].
 \label{eq:magnetic_conservation}
\end{equation}
The function $p(\sigma,a)$ is the surface equation of state. The radial equation describes the material response along the motion, with the bulk parameters and constitutive coefficients held fixed. It also extends smoothly through a turning point when obtained from the junction identity. There is no non-areal-area flux correction for Eq.~\eqref{eq:magnetic_metric}. Solving Eq.~\eqref{eq:magnetic_dynamic_density} for the radial velocity gives
\begin{equation}
 \dot a^2+V(a)=0.
 \label{eq:magnetic_motion}
\end{equation}
The dimensionless effective potential $V$ depends on the solution of Eq.~\eqref{eq:magnetic_conservation},
\begin{equation}
 V(a)=f(a)-4\pi^2a^2\sigma(a)^2.
 \label{eq:magnetic_potential}
\end{equation}
The negative density branch selected by the un-squared junction condition remains the physical branch of $V$. A static configuration requires $V(a_0)=V'(a_0)=0$; the stresses in Eq.~\eqref{eq:magnetic_static_stresses} ensure both equalities once the equation of state is compatible at $a_0$. For a differentiable equation of state, the two independent local responses are
\begin{equation}
 \zeta=\left(\frac{\partial p}{\partial\sigma}\right)_a,
 \qquad p_{,a}=\left(\frac{\partial p}{\partial a}\right)_\sigma.
 \label{eq:magnetic_response}
\end{equation}
Both $\zeta$ and $p_{,a}$ are dimensionless in the present reduction. The first changes the pressure with density at fixed radius; the second measures its explicit radial dependence at fixed density. The derivative with respect to physical radius at fixed physical density is $p_{,a}/\ell^2$. A constitutive slope of an exotic surface fluid is not automatically the squared sound speed of a causal volumetric medium~\cite{PoissonVisser1995}. Differentiating Eq.~\eqref{eq:magnetic_potential} along the conservation law, with $dp/da=\zeta\,d\sigma/da+p_{,a}$, gives
\begin{equation}
 \begin{split}
 V''(a)={}&f''(a)-8\pi^2\bigl[(\sigma+2p)^2\\
 &\quad+2\sigma(1+2\zeta)(\sigma+p)\bigr]\\
 &+16\pi^2a\sigma p_{,a}.
 \end{split}
 \label{eq:magnetic_curvature_stress}
\end{equation}
The last term is independent of the density response and disappears only for an EOS without explicit radius dependence. It cannot be incorporated into $\zeta$ while preserving the definition in Eq.~\eqref{eq:magnetic_response}. This distinction is the same one that underlies variable surface fluids in the Schwarzschild construction~\cite{Varela2015} and in related nonlinear-electrodynamic applications~\cite{Javed2023}. At equilibrium the same result takes a geometric form,
\begin{equation}
 \begin{split}
 V''(a_0)={}&f''(a_0)-\frac{f'(a_0)^2}{2f(a_0)}\\
 &+(1+2\zeta_0)\left[\frac{f'(a_0)}{a_0}-\frac{2f(a_0)}{a_0^2}\right]\\
 &-8\pi\sqrt{f(a_0)}\,p_{,a0}.
 \end{split}
 \label{eq:magnetic_curvature_geometry}
\end{equation}
The response values $\zeta_0$ and $p_{,a0}$ are evaluated at $(\sigma_0,a_0)$ after constitutive calibration. Equation~\eqref{eq:magnetic_curvature_geometry} retains the complete exponential geometry through compact lapse derivatives. Replacing it by its asymptotic charged approximation near the horizon would change both the equilibrium and its restoring response. A small radial displacement in the reduced proper time has the form
\begin{equation}
 a(u)=a_0+\epsilon\,\chi(u),\qquad |\epsilon|\ll1.
 \label{eq:magnetic_perturbation}
\end{equation}
The constant $\epsilon$ is a dimensionless perturbation amplitude and $\chi$ is a dimensionless radial mode. This is a displacement of one calibrated throat. Varying $a_0$ at fixed $\lambda$ instead labels separate equilibria, each with its own calibrated material coefficient. Along the perturbation, that coefficient, $\lambda$, and $\ell$ remain unchanged. The equilibrium conditions reduce the local expansion to
\begin{equation}
 V(a_0+\epsilon\chi)=\frac{\epsilon^2}{2}
 V''(a_0)\chi^2+\mathcal O(\epsilon^3).
 \label{eq:magnetic_potential_expansion}
\end{equation}
The corresponding linearized acceleration equation is
\begin{equation}
 \ddot\chi+\frac12 V''(a_0)\chi=0.
 \label{eq:magnetic_linear_mode}
\end{equation}
Thus $V''(a_0)>0$ gives an oscillatory radial mode, $V''(a_0)<0$ gives an exponentially growing mode, and $V''(a_0)=0$ is inconclusive at this order. The squared angular frequency measured in physical proper time is $V''(a_0)/(2\ell^2)$. Growth of the linear mode does not by itself establish collapse or a nonlinear final state. The constraint in Eq.~\eqref{eq:magnetic_motion} also restricts admissible initial data. At a positive-curvature minimum, holding exactly the same conservation integration constant gives $V=0$ only at equilibrium in a sufficiently small neighborhood. A finite excitation must satisfy the junction constraint and may change that integration constant at quadratic order while retaining the same constitutive coefficients. The linear acceleration criterion therefore concerns the response of compatible nearby data; it is not a claim that an arbitrary displacement can oscillate in a positive potential with all integration constants unchanged~\cite{Varela2015}.

\subsection{Reversible thermodynamics and constitutive thermal response}
\label{subsec:magnetic_thermodynamics}
The conservation law also admits a thermodynamic interpretation, following the framework of Lobo and Rodrigues~\cite{LoboRodrigues2026Thermodynamics}. Use the physical radius $r=\ell a$, area $A=4\pi r^2$, surface density $\sigma_{\rm phys}=\sigma/\ell$, and pressure $P=p/\ell$. The proper energy of the dynamic layer is $U=A\sigma_{\rm phys}=4\pi\ell a^2\sigma$; its static value is $E_{\rm shell}$ in Eq.~\eqref{eq:shell_proper_energy}, rather than the ADM mass. Equation~\eqref{eq:magnetic_conservation} gives

\begin{equation}
 \frac{dU}{d\tau}+P\frac{dA}{d\tau}=0.
 \label{eq:magnetic_thermo_balance}
\end{equation}
This balance holds for compatible radial motion, including turning points. The bulk is nonvacuum, but $\rho+p_r=0$ makes the energy flux into the shell vanish on each side. Transparency in this balance therefore does not require a vacuum exterior. Assume, additionally, reversible local equilibrium with entropy $S$, shell temperature $T_{\rm sh}>0$, fixed particle number and composition, and no other work variables. The first law then implies

\begin{equation}
 T_{\rm sh}\,dS=dU+P\,dA,
 \qquad \frac{dS}{d\tau}=0.
 \label{eq:magnetic_shell_isentropy}
\end{equation}

Thus all three constitutive families admit an isentropic interpretation under these assumptions, independently of the sign of $V''(a_0)$. This result concerns evolution with fixed bulk parameters and material coefficients; it does not assign a common entropy to separately calibrated equilibria or to scans that change the electromagnetic coupling. Isentropy does not imply constant temperature. The junction geometry supplies a distinct local acceleration scale, relevant to generalized Unruh constructions~\cite{LoboRodrigues2026Unruh}. Let $v_\Sigma^\mu$ be the physical unit four-velocity of the layer and $b_\Sigma^\mu=v_\Sigma^\nu\nabla_\nu v_\Sigma^\mu$ its acceleration evaluated from either identical bulk. With dots still denoting $d/du$, its magnitude is

\begin{equation}
 \alpha_\Sigma\equiv\sqrt{b_{\Sigma\mu}b_\Sigma^\mu}
 =\frac{|\ddot a+f'(a)/2|}
 {\ell\sqrt{f(a)+\dot a^2}}
 =2\pi|\sigma_{\rm phys}+2P|.
 \label{eq:magnetic_acceleration_stress}
\end{equation}

The last equality follows exactly from Eqs.~\eqref{eq:magnetic_dynamic_density} and~\eqref{eq:magnetic_dynamic_pressure}. Retaining the reduced Planck constant $\hbar$ and Boltzmann's constant $k_{\rm B}$, define the acceleration temperature scale

\begin{equation}
 T_{\rm acc}=\frac{\hbar\alpha_\Sigma}{2\pi k_{\rm B}}
 =\frac{\hbar}{k_{\rm B}\ell}|\sigma+2p|.
 \label{eq:magnetic_acceleration_temperature}
\end{equation}

This gives a geometric interpretation to the stress combination plotted in Fig.~\ref{fig:exponential-shell-pressure-energy}, without changing its distinction from the intrinsic SEC. It does not identify $T_{\rm acc}$ with the material temperature $T_{\rm sh}$ or with a particle-creation temperature. Such identifications require additional constitutive and quantum-state assumptions; equality with a peeling scale also requires the appropriate null-ray map and adiabatic regime~\cite{LoboRodrigues2026Stochastic}. For a static exterior throat,

\begin{equation}
 \begin{aligned}
 T_{\rm acc,0}&=\frac{\hbar f'(a_0)}
 {4\pi k_{\rm B}\ell\sqrt{f(a_0)}},\\
 \lim_{a_0\to\xi_+^+}\sqrt{f(a_0)}T_{\rm acc,0}
 &=\frac{\hbar f'(\xi_+)}{4\pi k_{\rm B}\ell}
 \equiv T_{\rm H}^{\rm seed}.
 \end{aligned}
 \label{eq:magnetic_acceleration_horizon_limit}
\end{equation}

At a fixed non-extremal seed, the local scale diverges while its redshifted limit equals the seed Hawking temperature. The retained wormhole has no horizon, and this equality alone establishes neither thermal emission nor an asymptotic Hawking flux. A thermodynamic completion requires a fundamental relation $U=U(S,A)$, with $T_{\rm sh}=(\partial U/\partial S)_A$ and $P=-(\partial U/\partial A)_S$. To obtain a conditional thermal response, suppose that the mechanical law $p(\sigma,a)$ extends to a differentiable pressure $P(U,A)$ in a neighborhood of the chosen isentrope, with the same constitutive coefficients and no additional explicit entropy dependence at fixed $(U,A)$. The conservation profiles in Sec.~\ref{sec:constitutive} then determine the area dependence, while an entropy-dependent integration function remains free. Equality of the mixed derivatives of $U$ gives

\begin{equation}
 \begin{aligned}
 \left(\frac{\partial T_{\rm sh}}{\partial A}\right)_S
 &=-\left(\frac{\partial P}{\partial S}\right)_A
 =-\frac{\zeta}{A}T_{\rm sh},\\
 \left(\frac{d\ln T_{\rm sh}}{d\ln A}\right)_S&=-\zeta.
 \end{aligned}
 \label{eq:magnetic_thermal_integrability}
\end{equation}

Here $(\partial P/\partial U)_A=\zeta/A$, and $\zeta$ retains the definition in Eq.~\eqref{eq:magnetic_response}. This Maxwell relation supplies the temperature variation under the stated extension, rather than an identification with $T_{\rm acc}$. Let $T_0>0$ and $U_0<0$ denote temperature and proper energy at the calibrated equilibrium. Integrating along its isentrope yields, for the three laws introduced in Sec.~\ref{sec:constitutive},

\begin{equation}
 \begin{aligned}
 \frac{T_{\rm B}(a)}{T_0}
 &=\left(\frac{a}{a_0}\right)^{-2w_s},\\
 \frac{T_{\rm P}(a)}{T_0}
 &=\exp\!\left[\frac{2A_{\rm P}}{n}
 \left(a^{-n}-a_0^{-n}\right)\right],\quad n\ne0,\\
 \frac{T_{\rm C}(a)}{T_0}
 &=\frac{U_0}{U(a)}
 =\frac{a_0^2\sigma_0}{a^2\sigma(a)}.
 \end{aligned}
 \label{eq:magnetic_constitutive_temperatures}
\end{equation}

The subscripts label the barotropic, variable phantom, and variable Chaplygin completions of $T_{\rm sh}$. Here $w_s$ is the dimensionless barotropic pressure-to-density ratio, while $A_{\rm P}$ and $n$ are the dimensionless phantom amplitude and radial exponent in the laws of Sec.~\ref{sec:constitutive}. The phantom expression has the continuous $n=0$ limit of the barotropic result with $w_s=A_{\rm P}$. The Chaplygin expression follows from $\zeta=-p/\sigma$ and $dU=-P\,dA$, and is valid wherever its negative-density branch remains regular. Since $w_s<0$ and $A_{\rm P}<0$, the first two temperatures increase under a small isentropic expansion; the Chaplygin temperature decreases because $-p/\sigma>0$. These are local responses of fixed materials, not derivatives along a sequence of recalibrated shells. Their normalization $T_0$ is not fixed by the junction conditions. The same thermodynamic extension defines an isentropic area modulus $K_S$, with dimensions of surface pressure,

\begin{equation}
 \begin{split}
 K_S&\equiv-A\left(\frac{\partial P}{\partial A}\right)_S\\
 &=\frac1\ell\left[\zeta(\sigma+p)-\frac{a}{2}p_{,a}\right].
 \end{split}
 \label{eq:magnetic_isentropic_modulus}
\end{equation}
This follows by differentiating the pressure along Eq.~\eqref{eq:magnetic_conservation}; positive $K_S$ means that expansion lowers the surface pressure on that isentrope. At the photon-sphere equilibrium of Sec.~\ref{subsec:photon_shell_stability}, $\sigma_0+p_0=0$ and both variable laws have $p_{,a0}=-np_0/a_0$, with physical equilibrium pressure $P_0=p_0/\ell$. Hence

\begin{equation}
 \begin{aligned}
 K_{S,\rm P}=K_{S,\rm C}&=\frac{nP_0}{2},\\
 \left.\frac{d\ln T_{\rm P}}{d\ln A}\right|_0&=+1,\qquad
 \left.\frac{d\ln T_{\rm C}}{d\ln A}\right|_0=-1,
 \end{aligned}
 \label{eq:magnetic_photon_thermal_response}
\end{equation}
The two materials can therefore have the same area modulus and radial restoring curvature at this radius, yet opposite temperature responses. Positive area stiffness for $n>0$ is insufficient for radial stability, which requires $n$ to exceed the critical exponent $n_{\rm ph}^{\rm crit}$ defined in Eq.~\eqref{eq:photon_common_threshold}. Neither stiffness nor the temperature ratios determine the heat capacity or the full thermodynamic stability: these depend on the remaining entropy dependence of $U$. Absolute entropy and temperature likewise remain unspecified. Dissipation, stochastic forcing, or absorbed radiation would require additional exchange physics beyond the present conservative and optically transparent shell.

\section{Constitutive dependence of the restoring response}\label{sec:constitutive}
The three surface laws below are imposed on the moving throat before their coefficients are fixed by the static junction. Each law selects a different pressure response for the same equilibrium stress pair. Their stability properties can consequently differ even though their densities and pressures agree at the chosen $a_0$.

\subsection{Linear barotropic fluid}\label{sec:linear_fluid}

A linear surface law provides a density-dependent reference without an additional radial factor,
\begin{equation}
 p=w_s\sigma.
 \label{eq:magnetic_barotropic_law}
\end{equation}
The constant $w_s$ is the ratio of pressure to density for this material and carries no units. Its mathematical domain is real. No universal numerical interval for $w_s$ follows from the phenomenological surface-fluid treatment~\cite{PoissonVisser1995,Varela2015}. In particular, a prior such as $0\leq w_s\leq1$ would presume a microscopic sound-speed interpretation not supplied by a negative-density surface law. We retain values compatible with the static junction in the admissible exterior. The boundary values $w_s=0$ and $w_s=-1$ describe, respectively, a pressureless law and a law saturating $\sigma+p=0$; neither algebraic limit guarantees equilibrium in the magnetic geometry. The local responses are $\zeta=w_s$ and $p_{,a}=0$, so all explicit radial dependence enters through the geometry and the density evolution. Static compatibility requires $p_0=w_s\sigma_0$, fixing
\begin{equation}
 w_s=-\frac12-\frac{a_0f'(a_0)}{4f(a_0)}.
 \label{eq:magnetic_barotropic_calibration}
\end{equation}
The coefficient is well defined for every strictly exterior throat, but cannot be calibrated on a horizon where $\sigma_0$ vanishes. Since $f'>0$ in this exterior, $w_s<-1/2$. The tangential NEC is equivalent to $w_s\leq-1$ at equilibrium, because multiplication by $\sigma_0<0$ reverses the inequality. This is a restriction on equilibrium matter, distinct from the radial stability criterion. With this fixed coefficient, the conservation law integrates to
\begin{equation}
 \sigma(a)=\sigma_0\left(\frac{a_0}{a}\right)^{2(1+w_s)}.
 \label{eq:magnetic_barotropic_density}
\end{equation}
The ratio of positive radii keeps the power real for any real $w_s$, and the solution preserves the negative density branch. The case $w_s=-1$ has constant density; $w_s=0$ dilutes as inverse area. Increasing $w_s$ changes this density response rather than the prescribed bulk mass. Differentiating the resulting potential or substituting the calibrated response into Eq.~\eqref{eq:magnetic_curvature_geometry} gives
\begin{equation}
 V_{\rm B}''(a_0)=f''(a_0)+\frac{f'(a_0)}{a_0}
 -\frac{f'(a_0)^2}{f(a_0)}.
 \label{eq:magnetic_barotropic_curvature}
\end{equation}
The subscript $\rm B$ identifies the barotropic specialization of the same potential. The last term is nonpositive and becomes dominant sufficiently close to a simple horizon. For any fixed non-extremal $\lambda$, $V_{\rm B}''\sim-f'(\xi_+)/(a_0-\xi_+)$ in that limit. Far from the throat scale, Eq.~\eqref{eq:magnetic_asymptotic} gives $V_{\rm B}''=-2\mu/a_0^3+\mathcal O(a_0^{-4})$, which is also negative. These limits establish instability near a simple horizon and at sufficiently large radius, without classifying an intermediate interval by extrapolation.

In the Schwarzschild limit at fixed physical mass, Eq.~\eqref{eq:magnetic_barotropic_curvature} reduces to $-2\mu/[a_0^2(a_0-2\mu)]$, negative for every $a_0>2\mu$. A general barotropic equation of state with a freely specified local slope is a different constitutive problem: here the full linear law fixes that slope to the calibrated pressure-to-density ratio.

\begin{figure}[tbp]
    \begin{minipage}{\columnwidth}
    \centering
    \includegraphics[width=\linewidth]{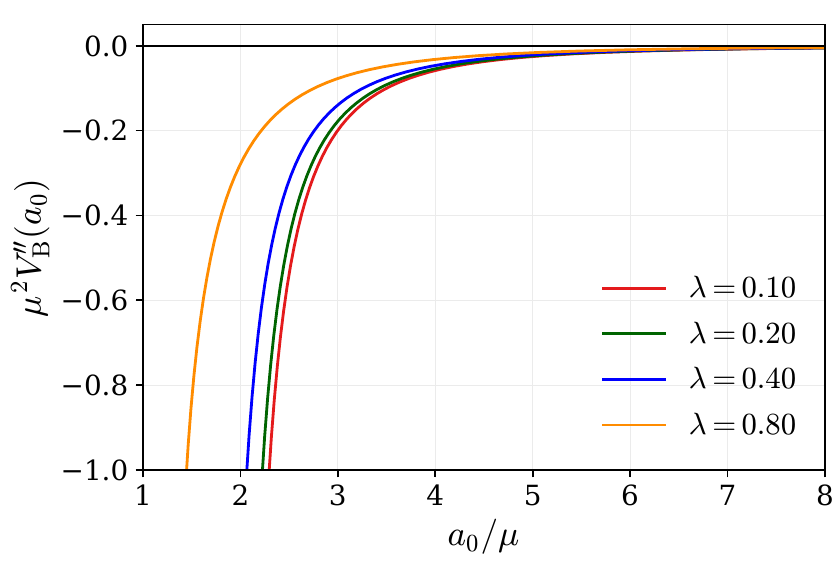}
    \caption{Mass-normalized radial stability curvature $\mu^2V_{\rm B}''(a_0)$ as a function of the equilibrium throat radius $a_0/\mu$ for the linear barotropic surface law in Eq.~\eqref{eq:magnetic_barotropic_law}. The four curves have $\lambda=0.10,0.20,0.40,0.80$ at fixed asymptotic mass $M$, with $\mu$ defined by Eq.~\eqref{eq:magnetic_mass}. The curvature is evaluated along the conservation law and checked against Eq.~\eqref{eq:magnetic_barotropic_curvature}. The coefficient $w_s$ is calibrated separately at each equilibrium using Eq.~\eqref{eq:magnetic_barotropic_calibration} and held fixed during its perturbation. Each curve includes only the static exterior $a_0>\xi_+(\lambda)$ satisfying Eq.~\eqref{eq:magnetic_field_domain}. The black horizontal line denotes zero; positive and negative curvatures correspond to stable and unstable linear radial modes, respectively. The negative nonextremal horizon divergences extend below the displayed vertical window.}
    \label{fig:exponential-barotropic-stability}
    \end{minipage}
\end{figure}

The barotropic curvature in Fig.~\ref{fig:exponential-barotropic-stability} remains negative along all four admissible branches in the displayed interval. Although each profile rises monotonically toward zero, no finite-radius crossing or positive-curvature window occurs in these configurations. According to Eq.~\eqref{eq:magnetic_linear_mode}, their radial response therefore contains a growing linear mode. At a common mass-normalized radius outside all four outer horizons, increasing $\lambda$ makes the curvature less negative. With $M$ fixed, this ordering corresponds to a lower growth rate measured in physical proper time, rather than a change to an oscillatory mode. The effect concerns the complete combination of lapse derivatives in Eq.~\eqref{eq:magnetic_barotropic_curvature}; the positive exterior lapse gradient alone does not ensure a restoring response. This finite family does not establish the sign for arbitrary geometric parameters, nor does its linear instability determine the subsequent nonlinear evolution.

The visible beginnings of the curves are set by the lower vertical plotting limit, not by a new material boundary. Each branch continues toward its own outer horizon with the negative divergence described above, while the throat remains strictly exterior. The inverse-lapse contribution in Eq.~\eqref{eq:magnetic_barotropic_curvature} accounts for this behavior. At larger radii, the decreasing curve separation is consistent with the common leading inverse-cubic decay in mass units obtained from Eq.~\eqref{eq:magnetic_asymptotic}. Approaching zero from below does not produce a marginal equilibrium at a finite radius. Comparison with Fig.~\ref{fig:exponential-shell-pressure-energy} also distinguishes the restoring response from the equilibrium energy inequalities: the near-horizon intervals with a positive tangential NEC combination still have negative barotropic curvature. Thus satisfaction of that inequality does not stabilize the calibrated linear surface law.

\subsection{Variable phantom surface fluid}\label{sec:variable_phantom}
We adopt the radius-dependent linear law studied by Varela~\cite{Varela2015},
\begin{equation}
 p=A_{\rm P}\,a^{-n}\sigma.
 \label{eq:magnetic_phantom_law}
\end{equation}
The amplitude $A_{\rm P}$ and exponent $n$ are dimensionless in the reduced variables. In physical units the corresponding amplitude is $\ell^nA_{\rm P}$, with dimensions of length to the power $n$. Both parameters have a real mathematical domain because $a>0$. Varela treats real exponents and identifies positive $n$ when examining outward-decaying factors and Schwarzschild stability sectors; those sectors are not universal restrictions on a magnetic seed. We allow the real local family, distinguishing the decaying subcase $n>0$ and the radius-independent boundary $n=0$. The designation phantom refers to the inherited functional form and does not impose a cosmological equation of state bound on the surface ratio.

At fixed amplitude, positive $n$ decreases the magnitude of the explicit factor outward, while negative $n$ increases it. The amplitude controls the sign and strength of the pressure to density ratio. The value $A_{\rm P}=0$ is pressureless and is incompatible with the positive static pressure of this exterior. The limit $n=0$ recovers Eq.~\eqref{eq:magnetic_barotropic_law} with $w_s=A_{\rm P}$. For finite positive radius the law has no pole in density.
Its two derivatives are
\begin{equation}
 \zeta=A_{\rm P}a^{-n},\qquad p_{,a}=-\frac{np}{a}.
 \label{eq:magnetic_phantom_response}
\end{equation}
The exponent modifies the explicit radial response independently of the density slope. Static compatibility $p_0=A_{\rm P}a_0^{-n}\sigma_0$ fixes
\begin{equation}
 A_{\rm P}=-\frac{a_0^n}{2}
 \left[1+\frac{a_0f'(a_0)}{2f(a_0)}\right].
 \label{eq:magnetic_phantom_calibration}
\end{equation}
The calibrated amplitude is negative on the retained exterior, so the positive pressure accompanies negative density. The exponent remains the free constitutive parameter across equilibria. Changing $n$ at one $a_0$ changes the material response while preserving the same equilibrium stresses by recalibrating the amplitude; the new amplitude then remains fixed during that material's perturbation. For $n\neq0$, the conserved density profile is
\begin{equation}
 \sigma(a)=\sigma_0\left(\frac{a_0}{a}\right)^2
 \exp\!\left[\frac{2A_{\rm P}}{n}
 \left(a^{-n}-a_0^{-n}\right)\right].
 \label{eq:magnetic_phantom_density}
\end{equation}
The exponential is positive, so the branch sign is preserved. As $n\to0$ at fixed amplitude, the exponent approaches $-2A_{\rm P}\ln(a/a_0)$ and the profile tends to Eq.~\eqref{eq:magnetic_barotropic_density}, rather than developing a physical singularity. The explicit radial contribution changes the potential curvature to
\begin{equation}
 V_{\rm P}''(a_0)=V_{\rm B}''(a_0)
 +n\frac{a_0f'(a_0)+2f(a_0)}{a_0^2}.
 \label{eq:magnetic_phantom_curvature}
\end{equation}
The subscript $\rm P$ denotes this variable phantom specialization. Because $a_0f'+2f>0$ in the exterior, increasing $n$ at fixed geometric equilibrium increases the local restoring curvature. Thus a positive exponent can compensate a negative barotropic response without changing the equilibrium energy-condition combinations. The exact local boundary is
\begin{equation}
 n_{\rm P}^{\rm crit}(a_0,\lambda)=
 -\frac{a_0^2V_{\rm B}''(a_0)}{a_0f'(a_0)+2f(a_0)}.
 \label{eq:magnetic_phantom_threshold}
\end{equation}
The critical exponent $n_{\rm P}^{\rm crit}$ separates positive and negative curvatures: stability requires $n>n_{\rm P}^{\rm crit}$. Equality is marginal at quadratic order. This threshold depends on both the throat radius and the magnetic shape parameter and is not a fundamental domain for $n$.

For every finite $n$, the extra term in Eq.~\eqref{eq:magnetic_phantom_curvature} stays finite at a simple horizon while the barotropic term diverges negatively. Therefore an arbitrarily close nonextremal throat remains unstable in this law. In contrast, its large-radius behavior is $V_{\rm P}''=2n/a_0^2+\mathcal O(a_0^{-3})$, so each fixed positive exponent gives positive curvature at sufficiently large radii within the separately calibrated family. Such a statement concerns distinct equilibria, not the evolution of one shell with a coefficient readjusted along its motion. At $n=0$ the barotropic curvature is recovered exactly. In the fixed-mass Schwarzschild limit the threshold becomes $\mu a_0/[(a_0-2\mu)(a_0-\mu)]$, reproducing the radius-dependent stabilization criterion of Ref.~\cite{Varela2015}.

\begin{figure*}[t]
    \begin{minipage}{\textwidth}
    \centering
    \includegraphics[width=\linewidth]{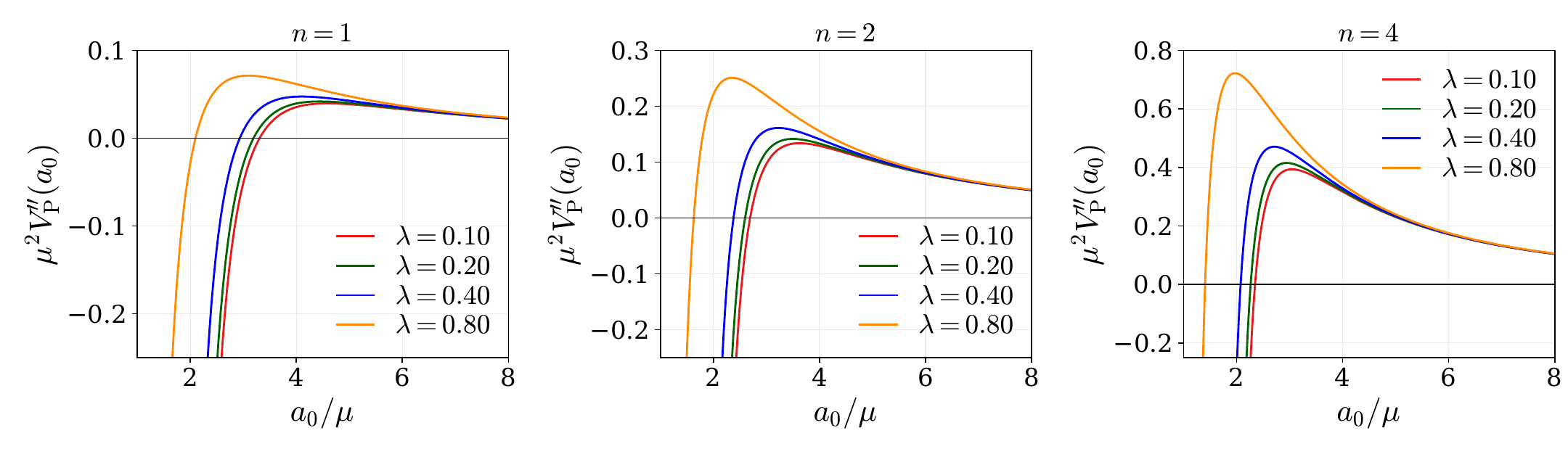}
    \caption{Mass-normalized radial stability curvature $\mu^2V_{\rm P}''(a_0)$ as a function of the equilibrium throat radius $a_0/\mu$ for the variable phantom surface law in Eq.~\eqref{eq:magnetic_phantom_law}. The three panels have $n=1,2,4$, from left to right. Each panel contains the four curves $\lambda=0.10,0.20,0.40,0.80$ at fixed asymptotic mass $M$, with $\mu$ defined by Eq.~\eqref{eq:magnetic_mass}. Vertical scales differ between panels. The curvature follows Eq.~\eqref{eq:magnetic_phantom_curvature}. The amplitude $A_{\rm P}$ is calibrated separately at each equilibrium by Eq.~\eqref{eq:magnetic_phantom_calibration} and held fixed, together with $n$, during that equilibrium's perturbation. Only the static exterior $a_0>\xi_+(\lambda)$ satisfying Eq.~\eqref{eq:magnetic_field_domain} is shown. The black horizontal line denotes zero; positive and negative curvatures correspond to stable and unstable linear radial modes, respectively. Negative non-extremal horizon divergences extend below the displayed vertical windows.}
    \label{fig:exponential-phantom-stability}
    \end{minipage}
\end{figure*}

Figure~\ref{fig:exponential-phantom-stability} exhibits a positive-curvature interval for each of the twelve configurations, in contrast to the negative barotropic profiles in Fig.~\ref{fig:exponential-barotropic-stability}. Each phantom curve crosses zero once in the displayed range, separating an inner unstable interval from equilibria with an oscillatory linear radial mode. For each selected exponent, increasing $\lambda$ through the four plotted values shifts this boundary inward in mass units. For example, the crossing moves from $a_0/\mu\simeq3.307$ to $2.099$ between $\lambda=0.10$ and $0.80$ when $n=1$, and from $2.348$ to $1.406$ when $n=4$. The zero itself requires higher-order analysis, as specified by Eq.~\eqref{eq:magnetic_linear_mode}. These values describe the sampled family rather than a bound for all allowed geometric parameters.

Increasing $n$ enlarges the positive curvature portion of the plotted exterior. The mechanism is the explicit pressure response in Eq.~\eqref{eq:magnetic_phantom_response}: its contribution to Eq.~\eqref{eq:magnetic_phantom_curvature} is positive because $a_0f'(a_0)+2f(a_0)>0$. At a common geometric equilibrium, the curvature therefore increases with $n$ even though the static junction stresses remain unchanged. The different vertical scales expose the resulting increase in the positive peaks. This constitutive stabilization leaves the negative surface density in Fig.~\ref{fig:exponential-shell-density} unchanged; it does not establish satisfaction of the energy conditions. The comparison between exponents involves separately calibrated amplitudes, each held fixed during the associated perturbation, rather than a coefficient adjusted along one shell's motion.

Beyond their positive maxima, the profiles decrease while remaining positive throughout the displayed outer interval. Their decreasing separation at fixed $n$ is consistent with the common leading large-radius behavior stated above; the curves need not retain a strict ordering in $\lambda$ after their crossings. Conversely, the negative branches continue below the lower plotting bounds toward their respective simple horizons. A finite exponent cannot remove that near-horizon instability, since the added radial-response term remains finite there. Thus the positive intervals identify local linear radial stability, without determining finite-amplitude dynamics or a nonlinear final state.

\subsection{Variable Chaplygin surface gas}\label{sec:variable_chaplygin}

The inverse-density law introduces a different constitutive slope for the same junction stresses~\revMagnetic{\cite{EiroaSimeone2007,EiroaFigueroa2012,Varela2015,Javed2023}},
\begin{equation}
 p=B_{\rm C}\frac{a^{-n}}{\sigma}.
 \label{eq:magnetic_chaplygin_law}
\end{equation}
The coefficient $B_{\rm C}$ is dimensionless here; in physical units its counterpart is $\ell^{n-2}B_{\rm C}$ and has dimensions of length to the power $n-2$. The exponent has a real mathematical domain, with the same interpretation of the explicit radial factor as in Eq.~\eqref{eq:magnetic_phantom_law}. Its value need not be an integer because the radius is positive. Unlike the phantom law, however, this equation of state requires $\sigma\neq0$ and has a density pole. The negative branch is regular in a sufficiently small neighborhood of any strictly exterior equilibrium, but cannot be calibrated directly at a horizon.

In the radius-independent limit $n=0$, a negative $B_{\rm C}$ reproduces the conventional Chaplygin sign $p=-|B_{\rm C}|/\sigma$. The radial extension in Ref.~\cite{Varela2015} allows real exponents; restrictions associated with a particular Schwarzschild stability window are results for that geometry rather than cosmological priors. We retain the real local surface family and distinguish $n>0$ as the explicitly decaying sector. Equilibrium compatibility determines the sign of the coefficient. This phenomenological surface relation does not inherit the volumetric density interpretation of a cosmological Chaplygin fluid. The responses are
\begin{equation}
 \zeta=-\frac{p}{\sigma},\qquad p_{,a}=-\frac{np}{a}.
 \label{eq:magnetic_chaplygin_response}
\end{equation}
At equilibrium the density slope has the opposite sign to that of the phantom law, even though their explicit-radius derivatives agree for the same $n$ and pressure. The static relation $p_0=B_{\rm C}a_0^{-n}/\sigma_0$ determines
\begin{equation}
 B_{\rm C}=-\frac{a_0^{n-2}}{16\pi^2}
 \left[a_0f'(a_0)+2f(a_0)\right].
 \label{eq:magnetic_chaplygin_calibration}
\end{equation}
Thus $B_{\rm C}<0$, preserving the conventional inverse-density sign on the retained exterior. Its absolute magnitude fixes the nonlinear pressure response and is not a second independent equilibrium variable. It is kept fixed with $n$ during the radial perturbation. The conservation law admits a closed expression valid on the negative branch,
\begin{equation}
 \begin{aligned}
 \sigma(a)^2={}&\frac1{a^4}\\
 &\times
 \begin{cases}
 \begin{aligned}
 &a_0^4\sigma_0^2\\[-0.2ex]
 &{}-\displaystyle\frac{4B_{\rm C}}{4-n}
 \left(a^{4-n}-a_0^{4-n}\right)
 \end{aligned},& n\neq4,\\[1.2ex]
 a_0^4\sigma_0^2-4B_{\rm C}\ln(a/a_0),& n=4.
 \end{cases}
 \end{aligned}
 \label{eq:magnetic_chaplygin_density}
\end{equation}
The density is the negative square root, defined where the bracket is strictly positive; its zeros terminate the regular inverse-density branch. The logarithmic case is the continuous $n\to4$ limit of the integrated expression, not an excluded equation of state. The apparent singularity of the integrated expression at this exponent is therefore removable.

After calibration the curvature becomes
\begin{equation}
 V_{\rm C}''(a_0)=f''(a_0)+(n+1)\frac{f'(a_0)}{a_0}
 +(2n-4)\frac{f(a_0)}{a_0^2}.
 \label{eq:magnetic_chaplygin_curvature}
\end{equation}
The subscript $\rm C$ identifies the Chaplygin specialization. The term proportional to $f'^2/f$ cancels because of the inverse-density slope. There is therefore no pressure-gradient divergence of that form in the local curvature, even though the equilibrium pressure itself diverges at a simple horizon. The expression is regular in $n$, including $n=0$ and $n=4$.

The exterior sign of $a_0f'+2f$ gives an explicit threshold,
\begin{equation}
 n_{\rm C}^{\rm crit}(a_0,\lambda)=
 \frac{4f(a_0)-a_0f'(a_0)-a_0^2f''(a_0)}
 {a_0f'(a_0)+2f(a_0)}.
 \label{eq:magnetic_chaplygin_threshold}
\end{equation}
Stability requires $n>n_{\rm C}^{\rm crit}$, where $n_{\rm C}^{\rm crit}$ is the local critical exponent of this family. At fixed throat, the derivative of $V_{\rm C}''$ with respect to $n$ equals that in the phantom law. Their offset nevertheless differs: subtraction of Eqs.~\eqref{eq:magnetic_phantom_curvature} and \eqref{eq:magnetic_chaplygin_curvature} gives $V_{\rm C}''-V_{\rm P}''=f'^2/f-4f/a_0^2$. The same geometric sign controlling the tangential NEC makes this difference nonnegative whenever Eq.~\eqref{eq:magnetic_tangent_nec} is nonnegative. In that radial sector the Chaplygin response is at least as restoring as the phantom response for equal exponent, without removing the negative density. For a fixed simple horizon the curvature has the finite exterior limit
\begin{equation}
 \lim_{a_0\to\xi_+^+}V_{\rm C}''(a_0)
 =f''(\xi_+)+(n+1)\frac{f'(\xi_+)}{\xi_+}.
 \label{eq:magnetic_chaplygin_horizon_limit}
\end{equation}
A positive value ensures a sufficiently small exterior interval with positive curvature, rather than an equilibrium on the horizon. Its sign depends on both $n$ and $\lambda$, so the simple-horizon behavior differs from the unavoidable negative divergence in the other two laws. As $\lambda\to\lambda_e$ from below, $f'(\xi_+)\to0$ while $f''(\xi_+)\to f''(\xi_e)>0$. Hence, for any fixed finite exponent, sufficiently near-extremal seeds have a positive Chaplygin curvature immediately outside their outer horizon. The allowed perturbation amplitude must still be smaller than the separation from that horizon, and a stable local radial mode does not regularize the divergent equilibrium pressure of a nonextremal near-horizon shell.

The large-radius expansion exposes another distinction,
\begin{equation}
 \begin{split}
 V_{\rm C}''(a_0)={}&\frac{2n-4}{a_0^2}+\frac{(6-2n)\mu}{a_0^3}\\
 &+\frac{16(n-8)}{5\lambda a_0^8}+\mathcal O(a_0^{-12}).
 \end{split}
 \label{eq:magnetic_chaplygin_asymptotic}
\end{equation}
The inverse-square magnetic-charge term in Eq.~\eqref{eq:magnetic_asymptotic} cancels from this linear differential combination. The leading explicitly exponential correction occurs at inverse eighth order; the mass still depends on $\lambda$ through Eq.~\eqref{eq:magnetic_mass}. For $n>2$ the far-exterior curvature is positive, and for $n<2$ it is negative. At $n=2$ the leading term is $2\mu/a_0^3>0$. These asymptotic signs do not fix the sign at intermediate radius or turn the density-pole restriction into a global existence statement.

In the fixed-mass Schwarzschild limit the nonlinear term disappears and Eq.~\eqref{eq:magnetic_chaplygin_curvature} reduces to $[(2n-4)a_0+(6-2n)\mu]/a_0^3$. Its exterior is unstable for $n\leq1$, has the finite stable interval $2\mu<a_0<\mu(3-n)/(2-n)$ for $1<n<2$, and is stable for all exterior radii when $n\geq2$, as found by Varela~\cite{Varela2015}. Those intervals provide a controlled limiting comparison. The complete exponential geometry instead obeys Eq.~\eqref{eq:magnetic_chaplygin_threshold}, and its near-extremal positive-curvature sector cannot be inferred from the Schwarzschild classification.

\begin{figure*}[t]
    \begin{minipage}{\textwidth}
    \centering
    \includegraphics[width=\linewidth]{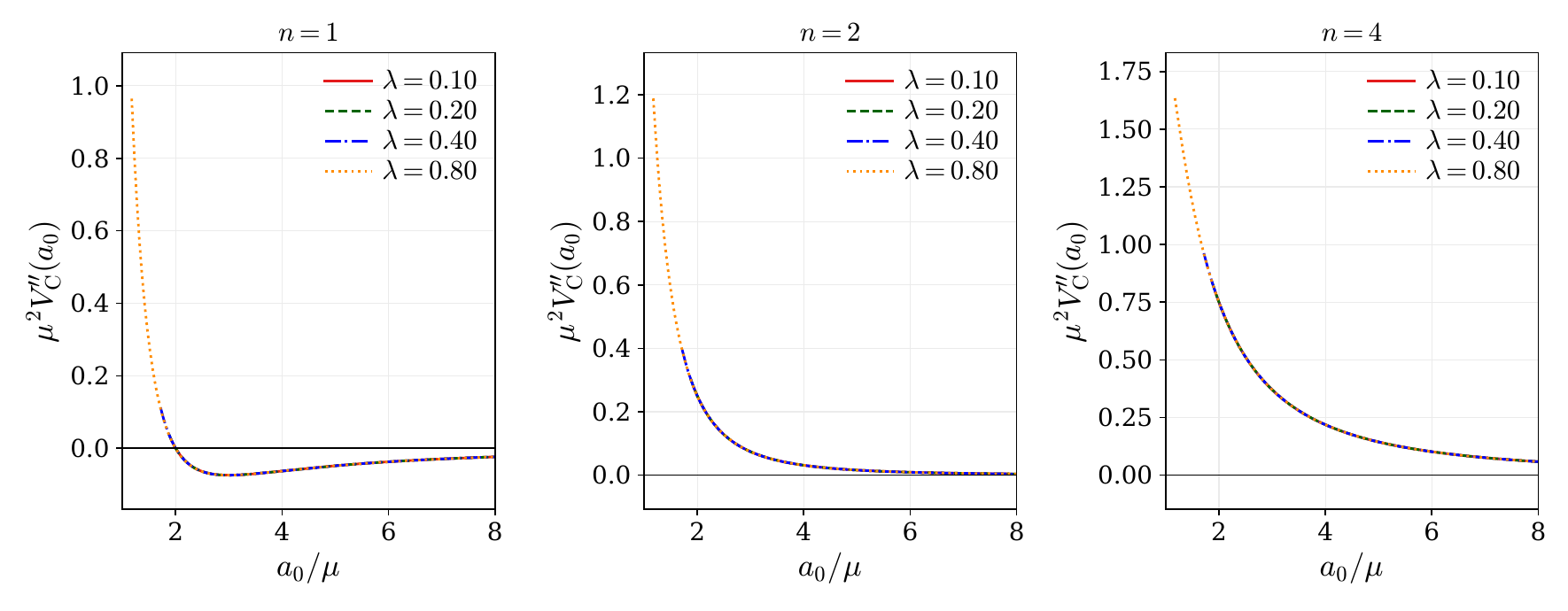}
    \caption{Mass-normalized radial stability curvature $\mu^2 V_{\rm C}''(a_0)$ as a function of $a_0/\mu$ for the variable Chaplygin law in Eq.~\eqref{eq:magnetic_chaplygin_law}. The panels have $n=1,2,4$ from left to right, and each contains $\lambda=0.10,0.20,0.40,0.80$ at fixed asymptotic mass $M$. The curves retain the red, green, blue and orange colors, respectively, and use solid, dashed, dash-dotted and dotted lines to distinguish nearly coincident profiles. Only $a_0>\xi_+(\lambda)$ is shown. The coefficient $B_{\rm C}$ is calibrated at each equilibrium by Eq.~\eqref{eq:magnetic_chaplygin_calibration} and held fixed during its perturbation; the curvature follows Eq.~\eqref{eq:magnetic_chaplygin_curvature}. Vertical scales differ between panels. The black horizontal line marks zero curvature; positive and negative values correspond to stable and unstable linear radial modes, respectively.}
    \label{fig:exponential-chaplygin-stability}
    \end{minipage}
\end{figure*}

The change of exponent in Fig.~\ref{fig:exponential-chaplygin-stability} separates a finite positive-curvature interval from the positive profiles found for the larger exponents. For $n=1$, each sampled geometry has a stable exterior interval adjacent to its outer horizon, followed by a sign change close to $a_0/\mu=2$ and a negative outer branch. The cases $n=2$ and $n=4$ remain positive throughout their displayed exterior domains. Their curves approach zero from above as the throat radius increases, consistently with the distinct leading orders in Eq.~\eqref{eq:magnetic_chaplygin_asymptotic}. The logarithmic conservation profile at $n=4$ therefore introduces no discontinuity into the restoring curvature.

The strong overlap of the four profiles at a common radius does not imply exact independence from $\lambda$. The curvature combination in Eq.~\eqref{eq:magnetic_chaplygin_curvature} cancels the inverse-square magnetic term of the lapse. At fixed physical mass, its leading large-radius terms consequently agree for all four geometries, while the first remaining nonlinear correction decays as the inverse eighth power of the radius. This suppression explains why the profiles become visually indistinguishable over much of the plotted interval. For example, at $a_0/\mu=2.5$ in the $n=4$ panel, $\mu^2V_{\rm C}''$ changes from approximately $0.5120000$ at $\lambda=0.10$ to $0.5119226$ at $\lambda=0.80$, a relative difference of about $0.015\%$. The nearly coincident curves, distinguished by their line styles, retain the small parameter dependence of the restoring response.

The parameter dependence is more evident in the accessible throat domain than in the overlapping outer curves. Across the sampled family, the lower boundary moves from $a_0/\mu\simeq1.93717$ at $\lambda=0.10$ to $1.17356$ at $\lambda=0.80$. In the $n=1$ panel, the zero shifts only from $a_0/\mu\simeq2.00000$ to $1.99675$, whereas the interval between the horizon and that zero widens substantially. The finite near-horizon curvature is compatible with Eq.~\eqref{eq:magnetic_chaplygin_horizon_limit}, but its positive sign does not remove the divergent equilibrium pressure at a simple horizon. Each point represents a separately calibrated equilibrium whose constitutive coefficient remains fixed during its own perturbation. These positive-curvature intervals establish local linear radial stability within the regular negative-density branch, without determining finite-amplitude motion or changing the sign of the surface density.

\subsection{Photon sphere equilibrium and explicit radius stabilization}
\label{subsec:photon_shell_stability}
The photon sphere condition also organizes the equilibrium matter and restoring response. Define $\mathcal U(\xi)=f(\xi)/\xi^2$ on the seed exterior, and let $\xi_{\rm ph}$ denote a nondegenerate unstable circular null orbit. It satisfies $\xi_{\rm ph}f'(\xi_{\rm ph})-2f(\xi_{\rm ph})=0$ and $\mathcal U''(\xi_{\rm ph})<0$. The same function will be denoted by $\mathcal U_{\rm ray}$ in the imaging calculation. For the equilibrium density of Eq.~\eqref{eq:magnetic_static_stresses}, direct differentiation yields

\begin{equation}
 \frac{d\sigma_0}{da_0}
 =-\frac{a_0f'(a_0)-2f(a_0)}
 {4\pi a_0^2\sqrt{f(a_0)}}
 =-\frac{2(\sigma_0+p_0)}{a_0}.
 \label{eq:density_nec_photon}
\end{equation}

Here the derivative labels the static equilibrium sequence; it does not recalibrate a material coefficient during a perturbation. At $a_0=\xi_{\rm ph}$ it vanishes, the tangential NEC is saturated, and $p_0=-\sigma_0=\sqrt{f(a_0)}/(2\pi a_0)$. In fact

\begin{equation}
 \left.\frac{d^2\sigma_0}{da_0^2}\right|_{\xi_{\rm ph}}
 =-\frac{\xi_{\rm ph}\mathcal U''(\xi_{\rm ph})}
 {4\pi\sqrt{f(\xi_{\rm ph})}}>0,
 \label{eq:density_min_photon}
\end{equation}
so the unstable photon sphere locates a strict local minimum of the static density. This explains the coincident numerical radii in the stress and orbital tables. For any differentiable $p(\sigma,a)$ calibrated at that radius, the density-slope term in Eq.~\eqref{eq:magnetic_curvature_geometry} vanishes. Since $f'=2f/a_0$ there, one obtains the exact identity

\begin{equation}
 \left.V''\right|_{a_0=\xi_{\rm ph}}
 =\xi_{\rm ph}^{\,2}\mathcal U''(\xi_{\rm ph})
 -8\pi\sqrt{f(\xi_{\rm ph})}\,p_{,a0}.
 \label{eq:photon_shell_identity}
\end{equation}
For a radius independent barotropic law, $p_{,a0}=0$, this curvature is strictly negative, independently of the finite local slope $\zeta_0$. This recovers the photon sphere stability obstruction for the transparent barotropic construction~\cite{TsukamotoKokubu2024}. It is stronger at this particular radius than a conclusion based only on the sampled calibrated linear laws. It does not assert instability at every other radius or for an arbitrary radius-dependent material. The two variable laws considered here both obey $p_{,a0}=-np_0/a_0$. Consequently they have identical restoring curvature at the photon sphere equilibrium, even though their density slopes differ

\begin{equation}
 \begin{split}
 V_{\rm P}''(\xi_{\rm ph})=V_{\rm C}''(\xi_{\rm ph})
 ={}&\xi_{\rm ph}^{\,2}\mathcal U''(\xi_{\rm ph})\\
 &+\frac{4nf(\xi_{\rm ph})}{\xi_{\rm ph}^{\,2}}.
 \end{split}
 \label{eq:photon_variable_equal}
\end{equation}
The common critical exponent is therefore
\begin{equation}
 n_{\rm ph}^{\rm crit}
 =\left.\frac{2f-\xi^2f''}{4f}\right|_{\xi_{\rm ph}}>0,
 \qquad n>n_{\rm ph}^{\rm crit}\quad\hbox{for stability}.
 \label{eq:photon_common_threshold}
\end{equation}

Equality is inconclusive at linear order. The positive explicit-radius response can overcome the negative optical-potential curvature without changing the equilibrium stress pair. For $\lambda=0.10,0.20,0.40,0.80$, respectively, the exact lapse gives $n_{\rm ph}^{\rm crit}\simeq1.4782,1.4529,1.3864,1.0517$; the fixed mass Schwarzschild limit gives $3/2$. These are thresholds at the seed photon-sphere radius, which is an exterior timelike throat radius, rather than at a seed horizon. The result ties the material response to the optical transition without identifying shell oscillations with photon orbit perturbations.

\subsection{Radial support of the imaged throat configurations}
\label{subsec:imaged_stability}
The optical comparisons in Sec.~\ref{sec:optical_appearance} use $a_0/\xi_+=1.2$ and $2$ for $\lambda=0.10,0.40,0.80$. Evaluating the already derived thresholds at those same radii connects the optical maps to a specified local material response. Table~\ref{tab:imaged_stability_thresholds} lists the results. Each entry uses its own static calibration, held fixed in the subsequent perturbation.

\begin{table}[tbp]

\caption{Critical exponents at the throat radii used in the images. Stability requires $n>n_{\rm P}^{\rm crit}$ or $n>n_{\rm C}^{\rm crit}$ for the corresponding material. The final column indicates whether the throat lies below the seed photon sphere; all six radii lie below the corresponding ISCO. Values use the exact exponential lapse.}
\label{tab:imaged_stability_thresholds}
\centering
\begin{tabular}{ccrrc}
\hline
$\lambda$ & $a_0/\xi_+$ & $n_{\rm P}^{\rm crit}$ & $n_{\rm C}^{\rm crit}$ & $a_0<\xi_{\rm ph}$\\
\hline
0.10 & 1.2 & 4.2728 & 1.2451 & Yes\\
0.40 & 1.2 & 4.2160 & 1.0563 & Yes\\
0.80 & 1.2 & 3.9579 & $-0.3950$ & Yes\\
0.10 & 2.0 & 0.6686 & 1.6521 & No\\
0.40 & 2.0 & 0.6790 & 1.5889 & No\\
0.80 & 2.0 & 0.7955 & 1.2587 & No\\
\hline
\end{tabular}
\end{table}
For the smaller throats, variable Chaplygin matter with $n=2$ or $n=4$ gives positive curvature in every imaged column. At $n=1$ it does so only for $\lambda=0.80$. Variable phantom matter with $n=1$ or $n=2$ is unstable at each smaller throat; for $n=4$, only the $\lambda=0.80$ smaller throat is stable. The proximity of $n=4$ to its last threshold is a quantitative result, not a stability classification inferred from image morphology. For the larger throats, each of the plotted phantom exponents $n=1,2,4$ exceeds the corresponding threshold. Chaplygin support is unstable at $n=1$ and stable at $n=2,4$. Direct evaluation of Eq.~\eqref{eq:magnetic_barotropic_curvature} gives negative curvature for all six calibrated linear barotropes. Thus one stationary optical map may have stable or unstable radial realizations depending on the constitutive law. These classifications concern local radial modes only.

In particular, the smaller-throat black-hole image degeneracy described below is compatible with stable radial Chaplygin support at $n=2$ and $n=4$ in all three geometries. At those radii the seed photon sphere survives the excision. The larger throats remove it but can still have stable radial support under the exponents just identified. The distinction between the two optical sectors therefore does not coincide with a universal transition from stable to unstable shell matter.

\section{Optical appearance of the static magnetic wormhole}\label{sec:optical_appearance}

\subsection{Thin-disk model}\label{subsec:optical_model}

We consider a geometrically thin, transparent emitting disk in the equatorial plane of the static wormhole. Its matter is neutral and follows circular timelike geodesics; self gravity, magnetic forces and radiation feedback are neglected. The emitted light is an independent, minimally coupled test field in the geometric-optics limit and propagates along null geodesics of Eq.~\eqref{eq:magnetic_metric}. This prescription does not describe disturbances of the nonlinear electromagnetic field supporting the geometry, whose characteristic cone can differ from the gravitational one~\cite{Novello2000,MurkSoranidis2024}. The images are therefore conditional predictions for this probe sector. Identifying the imaged radiation with the same electromagnetic field that supports the seed would require a separate characteristic and transfer analysis. The throat remains at $a_0$ and is transparent to this radiation, with no absorption, reflection or imposed refractive response.
A circular emitter has four-velocity $U_{\rm em}^{\alpha}=U_{\rm em}^{T}(1,0,0,\Omega)$, where $\Omega=d\phi/dT$. Timelike normalization and the radial geodesic equation give
\begin{equation}
 \begin{aligned}
 \Omega^2(\xi)&=\frac{f'(\xi)}{2\xi},\\
 \ell U_{\rm em}^{T}&=
 \left[f(\xi)-\frac{\xi f'(\xi)}2\right]^{-1/2}.
 \end{aligned}
 \label{eq:magnetic_opt_circular_velocity}
\end{equation}
The positive root of $\Omega$ fixes the disk rotation, and its physical angular velocity is $\Omega/\ell$. Circular emitters require $f'>0$ and $2f-\xi f'>0$. The inner boundary of the stable timelike branch connected to infinity is the ISCO, denoted by $\xi_{\rm I}$, which satisfies
\begin{equation}
 \left[\xi f f''+3f f'-2\xi(f')^2\right]_{\xi=\xi_{\rm I}}=0.
 \label{eq:magnetic_opt_isco}
\end{equation}
On the admitted circular branch, positivity of this bracket gives radial stability at fixed particle energy and angular momentum. We place the luminous inner edge at $\xi_{\rm in}=\xi_{\rm I}>a_0$ and the outer edge at $\xi_{\rm out}=1000\mu$, corresponding to $r_{\rm out}=1000M$. There is no emitting plunging region. A throat that excises the ISCO would require a different retained emitter support and is outside this disk prescription. An edge centered Gralla-Lupsasca-Marrone (GLM) profile~\cite{GrallaLupsascaMarrone2020} supplies the radial surface emission, following the moving-emitter construction of Ref.~\cite{Zhong2026Optical}. In the reduced variables, its unit-peak form is
\begin{equation}
 \begin{split}
 \mathcal I_{\rm em}(\xi)={}&
 \frac{\mu\exp\!\left[-\frac12
  \operatorname{arsinh}^2\!\left(\frac{\xi-\xi_{\rm in}}{\mu}\right)\right]}
 {\sqrt{(\xi-\xi_{\rm in})^2+\mu^2}}\\
 &\times H(\xi-\xi_{\rm in})H(\xi_{\rm out}-\xi).
 \end{split}
 \label{eq:magnetic_opt_emissivity}
\end{equation}
The step function $H$ includes the edge, and the width is $M$ in physical units. The factor $\mu$ preserves a common peak when $\lambda$ changes at fixed mass. This phenomenological surface law specifies neither an accretion rate nor a zero-torque viscous boundary; its nonzero ISCO emission is an adopted source property.

For exterior $\varsigma\in\{+,-\}$, the emitted specific intensity at local photon frequency $\nu$ is
\begin{equation}
 I_{\nu,\rm em}^{(\varsigma)}(\xi)=
 q_{\rm em}^{(\varsigma)}I_*\mathcal I_{\rm em}(\xi),
 \qquad q_{\rm em}^{(\varsigma)}\in\{0,1\}.
 \label{eq:magnetic_opt_source_switch}
\end{equation}
The common scale $I_*>0$ and source switches define observer-side emission, $(q_{\rm em}^{(+)},q_{\rm em}^{(-)})=(1,0)$, or identical bilateral disks, $(1,1)$. The surface emission is isotropic in the matter frame and locally frequency-flat over the finite emitted band sampled by the rays. Each supported crossing contributes once, without self-occultation or a path-length multiplier. These assumptions describe a transparent surface source; they do not supply a bolometric luminosity or a finite-thickness transfer law. The observer inclination $\iota$ to the disk axis obeys $0\leq\iota<\pi/2$, excluding tangential propagation within an exactly edge-on, infinitesimally thin disk.

\subsection{Imaging method}\label{subsec:optical_imaging_method}

Rays are traced backward from a static observer in the $+$ exterior. For the future-directed phase covector $k_\alpha$, the conserved energy, axial angular momentum and squared total angular momentum are $\mathcal E=-k_T>0$, $\mathcal J_\phi=k_\phi$ and $\mathcal J^2=k_\theta^2+\mathcal J_\phi^2/\sin^2\theta$, respectively. The reduced impact parameters are $\mathfrak b=\mathcal J/\mathcal E\geq0$ and $\mathfrak b_\phi=\mathcal J_\phi/\mathcal E$, with $|\mathfrak b_\phi|\leq\mathfrak b$; their physical counterparts are multiplied by $\ell$. The null Hamiltonian of the dimensionless conformal metric $\bar g_{\alpha\beta}=g_{\alpha\beta}/\ell^2$ yields
\begin{equation}
 \begin{aligned}
 \left(\frac{d\xi}{d\kappa}\right)^2
    &=\mathcal E^2\left[1-\mathfrak b^2
                      \mathcal U_{\rm ray}(\xi)\right],\\
 \mathcal U_{\rm ray}(\xi)&=\frac{f(\xi)}{\xi^2},\qquad
 \left|\frac{d\varphi}{d\kappa}\right|=\frac{\mathcal J}{\xi^2}.
 \end{aligned}
 \label{eq:magnetic_opt_radial}
\end{equation}
Here $\kappa$ is an affine parameter and $\varphi$ is the accumulated orbital-plane angle. The geometric parameter $\lambda$ is fixed along each trajectory. The null potential differs from the shell potential in Eq.~\eqref{eq:magnetic_potential}. On an allowed monotonic radial leg between endpoint radii $\xi_1$ and $\xi_2$, the angular sweep is
\begin{equation}
 \mathcal A_{\rm ray}(\xi_1,\xi_2;\mathfrak b)=
 \int_{\min(\xi_1,\xi_2)}^{\max(\xi_1,\xi_2)}
 \frac{\mathfrak b\,d\xi}
 {\xi^2\sqrt{1-\mathfrak b^2 f(\xi)/\xi^2}}.
 \label{eq:magnetic_opt_angle_integral}
\end{equation}
The outer turning root reached from the observer, $\xi_{\rm t}$, satisfies $\mathfrak b^2 f(\xi_{\rm t})=\xi_{\rm t}^2$ and lies in the retained exterior. The radial integrals extend over connected allowed intervals where the square root is real. A simple turning endpoint is regularized by $\xi=\xi_{\rm t}+w_{\rm t}^2$, with auxiliary variable $w_{\rm t}\geq0$; a critical circular orbit is instead treated by its one-sided limit.

At the static symmetric junction, continuity of local frequency and tangential momentum gives
\begin{equation}
 \mathcal E_+=\mathcal E_-,\qquad
 \mathcal J_+=\mathcal J_-,\qquad
 \mathcal J_{\phi,+}=\mathcal J_{\phi,-}.
 \label{eq:magnetic_opt_shell_constants}
\end{equation}
The outward radial chart component reverses sign across the throat, whereas the tangential components do not. The identical lapse and clock normalization give unit local frequency transfer at the junction. The same angular coordinates identify the two exteriors, as in Sec.~\ref{sec:magnetic_throat}. The static observer is also a zero-angular-momentum observer (ZAMO): in this static diagonal metric, the two frames coincide. It lies at radius $\xi_{\rm o}$, polar angle $\iota$ and azimuth zero, where $f_{\rm o}=f(\xi_{\rm o})$. Its orthonormal spatial axes point along increasing $\xi$, $\theta$ and $\phi$. With inward sky separation $\Theta$ and screen azimuth $\psi$, the backward launched unit spatial direction is
\begin{equation}
 \left(n^{\hat\xi},n^{\hat\theta},n^{\hat\phi}\right)
 =\left(-\cos\Theta,\ \sin\Theta\sin\psi,
                       \sin\Theta\cos\psi\right).
 \label{eq:magnetic_opt_camera_direction}
\end{equation}
Thus positive vertical angle points along $+e_{\hat\theta}$: the camera uses the positive launched-ray convention, and the displayed axes follow this tetrad directly. For unit stereographic focal length,
\begin{equation}
 \begin{aligned}
 x_{\rm scr}&=2\tan(\Theta/2)\cos\psi,\\
 y_{\rm scr}&=2\tan(\Theta/2)\sin\psi,\qquad
 s_{\rm scr}=\sqrt{x_{\rm scr}^2+y_{\rm scr}^2}.
 \end{aligned}
 \label{eq:magnetic_opt_screen}
\end{equation}
The associated impact parameters are
\begin{equation}
 \begin{aligned}
 \mathfrak b&=\frac{\xi_{\rm o}}{\sqrt{f_{\rm o}}}
       \sin\!\left[2\arctan(s_{\rm scr}/2)\right],\\
 \mathfrak b_\phi&=-\mathfrak b\sin\iota\cos\psi.
 \end{aligned}
 \label{eq:magnetic_opt_screen_impact}
\end{equation}
The second sign uses future photon momentum, opposite to the backward launch tangent. It fixes the approaching side of the rotating disk independently of the source palette. The observer radius is $\xi_{\rm o}=200\mu$. The Minkowski reference has a full angular field of $120^\circ$ per axis, with square half-width $2\tan(120^\circ/4)$. The wormhole images resolve its central $6^\circ$ field on each axis; this restriction changes the displayed angular range without changing the ray dynamics.

For the inclined-disk images in Figs.~\ref{fig:magnetic-inclined-disk} and~\ref{fig:magnetic-bilateral-disk}, the vertical display axis is reversed relative to Eq.~\eqref{eq:magnetic_opt_camera_direction}, so that its positive direction is $-e_{\hat\theta}$. The horizontal axis, disk rotation and observer colatitude are unchanged; the celestial images retain the original orientation.

The retained null potential determines the boundary between returning and transmitted directions. Its outer unstable circular orbit obeys
\begin{equation}
 \begin{gathered}
 \xi_{\rm ph}f'(\xi_{\rm ph})-2f(\xi_{\rm ph})=0,\\
 \mathcal U_{\rm ray}''(\xi_{\rm ph})<0,\qquad
 \mathfrak b_{\rm ph}=
 \frac{\xi_{\rm ph}}{\sqrt{f(\xi_{\rm ph})}}.
 \end{gathered}
 \label{eq:magnetic_opt_photon_sphere}
\end{equation}
For an observer outside the potential maximum, the transmission threshold is
\begin{equation}
 \mathfrak b_{\rm crit}^{-2}=
 \max_{\xi\geq a_0}\mathcal U_{\rm ray}(\xi).
 \label{eq:magnetic_opt_global_threshold}
\end{equation}
The four geometries considered here have a single exterior bulk maximum, giving
\begin{equation}
 \mathfrak b_{\rm crit}=
 \begin{cases}
 \mathfrak b_{\rm ph},&a_0\leq\xi_{\rm ph},\\[1ex]
 a_0/\sqrt{f(a_0)},&a_0>\xi_{\rm ph}.
 \end{cases}
 \label{eq:magnetic_opt_threshold_branches}
\end{equation}
Its angular radius follows directly from Eq.~\eqref{eq:magnetic_opt_screen_impact}, consistent with the finite-distance static-observer construction~\cite{PerlickTsupko2022}, through $\sin\Theta_{\rm crit}=\mathfrak b_{\rm crit}\sqrt{f_{\rm o}}/\xi_{\rm o}$.

\begin{table}[tbp]
\centering
\caption{Mass-normalized characteristic radii and null impact parameters for the exact magnetic lapse. The seed photon sphere contributes to the wormhole image only when it survives the excision at $a_0$.}
\label{tab:magnetic-optical-scales}
\begin{tabular}{ccccc}
\hline
$\lambda$ & $\xi_+/\mu$ & $\xi_{\rm ph}/\mu$ & $\xi_{\rm I}/\mu$ & $\mathfrak b_{\rm ph}/\mu$\\
\hline
0.10 & 1.937167 & 2.916532 & 5.813312 & 5.088117\\
0.20 & 1.869806 & 2.827827 & 5.617491 & 4.974240\\
0.40 & 1.716346 & 2.629721 & 5.191073 & 4.724135\\
0.80 & 1.173557 & 2.051512 & 4.080292 & 4.051679\\
\hline
\end{tabular}
\end{table}

Table~\ref{tab:magnetic-optical-scales} gives the exact-lapse orbital scales for $\lambda=0.10,0.20,0.40,0.80$, within the physical range established in Sec.~\ref{sec:exponential}. At fixed $M$, increasing $\lambda$ along this family decreases both the disk inner-edge radius and the seed critical impact parameter. The latter falls from $5.088117M$ to $4.051679M$ in physical units. This contracts the critical angular boundary when the light ring is retained, at the common observer radius; it does not determine a monotonic intensity or image-peak trend. The fixed-mass Schwarzschild limit recovers $r_{\rm ph}/M=3$, $r_{\rm I}/M=6$ and $b_{\rm ph}/M=3\sqrt3$, using the mass relation in Eq.~\eqref{eq:magnetic_mass} rather than substituting $\lambda=0$ into the reduced lapse.

The diagnostic celestial spheres lie at $\xi_{\rm s}=1000\mu$. Returning and transmitted rays reach these finite boundaries with angular sweeps
\begin{equation}
 \begin{aligned}
 \varphi_{\rm ret}&=
 \mathcal A_{\rm ray}(\xi_{\rm t},\xi_{\rm o};\mathfrak b)
 +\mathcal A_{\rm ray}(\xi_{\rm t},\xi_{\rm s};\mathfrak b),\\
 \varphi_{\rm tr}&=
 \mathcal A_{\rm ray}(a_0,\xi_{\rm o};\mathfrak b)
 +\mathcal A_{\rm ray}(a_0,\xi_{\rm s};\mathfrak b).
 \end{aligned}
 \label{eq:magnetic_opt_source_angles}
\end{equation}
Writing $\boldsymbol e_{\rm o}$ for the observer radial unit vector and $\boldsymbol e_{\theta,\rm o},\boldsymbol e_{\phi,\rm o}$ for its tangent basis, the source direction is
\begin{equation}
 \begin{split}
 \boldsymbol N_{\rm s}={}&\cos\varphi\,\boldsymbol e_{\rm o}\\
 &+\sin\varphi\left(\cos\psi\,\boldsymbol e_{\phi,\rm o}
              +\sin\psi\,\boldsymbol e_{\theta,\rm o}\right).
 \end{split}
 \label{eq:magnetic_opt_source_endpoint}
\end{equation}
The celestial texture comprises two orthogonal families of latitude circles, defined by $\arcsin(\boldsymbol N_{\rm s}\cdot\boldsymbol e_{\phi,\rm o})$ and $\arcsin(\boldsymbol N_{\rm s}\cdot\boldsymbol e_{\theta,\rm o})$. Black bands occupy angular distances within $0.006\,\mathrm{rad}$ of the latitude levels spaced by $10^\circ$. The signs of the two tangent components divide the sphere into four colored sectors, giving golden yellow, coral, ivory and wine clockwise from the upper left in the equatorial flat view. The source pattern is fixed in this observer-aligned frame for each image; varying inclination rotates its axes with the observer basis. The disk plane remains the fixed equatorial plane. The colors label source positions independently of the disk specific intensity.

\begin{figure}[tbp]
    \begin{minipage}{\columnwidth}
    \centering
    \includegraphics[width=\linewidth]{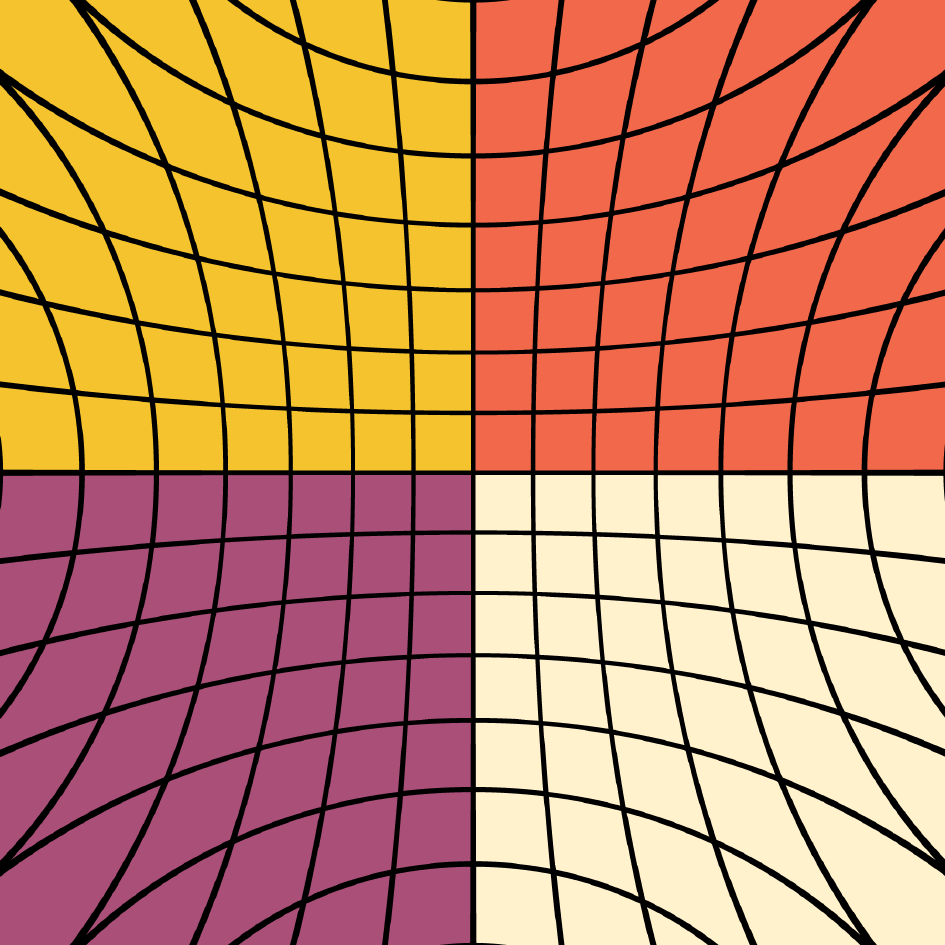}
    \caption{Celestial grid in Minkowski spacetime, viewed by an equatorial observer at radius $200M$ inside a source sphere of radius $1000M$. The stereographic field of view is $120^\circ$ along each screen axis. The two orthogonal source-latitude families have $10^\circ$ spacing and black bands with angular half-width $0.006\,\mathrm{rad}$. Clockwise from the upper left, the sectors are golden yellow, coral, ivory and wine. The scale $M$ is a reference length in this flat geometry, which contains no throat.}
    \label{fig:exponential-minkowski-grid}
    \end{minipage}
\end{figure}

The curved grid lines in Fig.~\ref{fig:exponential-minkowski-grid} arise even though all rays are straight in Minkowski spacetime. Each ray intersects a finite spherical source, and Eq.~\eqref{eq:magnetic_opt_screen} projects its direction onto the stereographic screen. The resulting latitude contours bend progressively away from the center; this curvature reflects the source and projection geometry rather than gravitational deflection. The grid retains reflection symmetry about both screen axes, as follows from the spherical boundary and the two equally spaced, orthogonal latitude families.

The central black cross consists of the zero-latitude bands of the two source families. It carries the same angular-band prescription as the remaining grid and does not bound a capture or transmission region. The four colors distinguish source sectors, without encoding frequency shifts or transported disk intensity. This flat image separates the angular structure of the illumination from the returning and transmitted endpoint maps in Eq.~\eqref{eq:magnetic_opt_source_angles}; changes in those maps alter the positions of a fixed source grid on the observer's sky.

Disk images use every supported equatorial crossing of each ray. For the observer four-velocity $U_{\rm o}^{T}=1/(\ell\sqrt{f_{\rm o}})$, the measured frequency ratio is $\mathfrak g=(-k_\alpha U_{\rm o}^{\alpha})/(-k_\alpha U_{\rm em}^{\alpha})$. Equations~\eqref{eq:magnetic_opt_circular_velocity} and \eqref{eq:magnetic_opt_shell_constants} then give
\begin{equation}
 \mathfrak g(\xi,\mathfrak b_\phi)=
 \frac{\sqrt{f(\xi)-\xi f'(\xi)/2}}
      {\sqrt{f_{\rm o}}\,[1-\Omega(\xi)\mathfrak b_\phi]}.
 \label{eq:magnetic_opt_frequency_circular}
\end{equation}
The denominator is positive for future-directed emission. Conservation of $I_\nu/\nu^3$ between crossings \revMagnetic{\cite{Younsi2012}} yields the specific intensity at observer frequency $\nu_{\rm o}$
\begin{equation}
 I_{\nu,\rm o}(x_{\rm scr},y_{\rm scr})=
 \sum_{j_{\rm d}}\mathfrak g_{j_{\rm d}}^3
 I_{\nu_{\rm o}/\mathfrak g_{j_{\rm d}},\rm em}^{(\varsigma_{j_{\rm d}})}
       (\xi_{j_{\rm d}}).
 \label{eq:magnetic_opt_intensity_sum}
\end{equation}
The index $j_{\rm d}$ enumerates supported disk crossings and records their radius and exterior; shared endpoints of adjacent radial legs are counted once. The cubic transfer concerns fixed observer frequency, whereas a bolometric fourth-power law requires a separately specified finite spectral integral. The critical-direction integrals are defined by one-sided limits, and finite crossing and pixel resolution limits the accuracy of narrow intensity peaks.

For display, each explicitly defined comparison set uses a single $I_{\rm ref}>0$, equal to its maximum sampled raw intensity over the cases and pixels in that set. Figure~\ref{fig:magnetic-inclined-disk} and Fig.~\ref{fig:magnetic-bilateral-disk} use separate comparison sets, as specified in their captions. For nonnegative raw specific intensity $I$, the common logarithmic map is
\begin{equation}
 \mathcal D(I)=
 \begin{cases}
 [\log_2(1+I_{\rm ref}/I)]^{-1},&I>0,\\
 0,&I=0.
 \end{cases}
 \label{eq:magnetic_opt_display}
\end{equation}
Transport and source side addition precede this transformation. Within each comparison set, the common reference preserves differences between panel maxima. The dimensionless display value represents the transformed intensity rather than physical radiance. The characteristic radii in Table~\ref{tab:magnetic-optical-scales} were independently checked by bracketing the exterior roots of Eqs.~\eqref{eq:magnetic_horizon}, \eqref{eq:magnetic_opt_photon_sphere} and \eqref{eq:magnetic_opt_isco}, with an absolute root tolerance of $10^{-12}$ in $\xi$. The exact-lapse first and second derivatives were used throughout; the charged asymptotic expansion was not substituted into these root equations. The resulting radii and critical impact parameters agree with the displayed table at its quoted precision. The quoted celestial critical diameters were also checked directly from Eqs.~\eqref{eq:magnetic_opt_screen_impact} and \eqref{eq:magnetic_opt_threshold_branches}. These checks validate the orbital and angular reference scales, rather than establishing convergence of the image intensities.
The images are finite-resolution realizations of the crossing prescription in Eq.~\eqref{eq:magnetic_opt_intensity_sum}. Unresolved near-critical structure can affect sampled peak intensities and the count of faint supported pixels. Accordingly, the image percentages below are approximate diagnostics of the stated screen and source prescription; no convergence-certified accuracy is assigned to them. A sum over uniformly spaced stereographic pixels is also not an integral over equal solid angles. For the stereographic coordinates used here, $d\Omega=dx_{\rm scr}\,dy_{\rm scr}/[1+s_{\rm scr}^2/4]^2$. We retain unweighted pixel sums only as finite-screen diagnostics and do not identify them with observed fluxes or total luminosities.

\FloatBarrier
\subsection{Imaging results}\label{subsec:optical_imaging_results}

\begin{figure*}[t]
    \noindent\begin{minipage}{\textwidth}
    \centering
    \includegraphics[width=\linewidth]{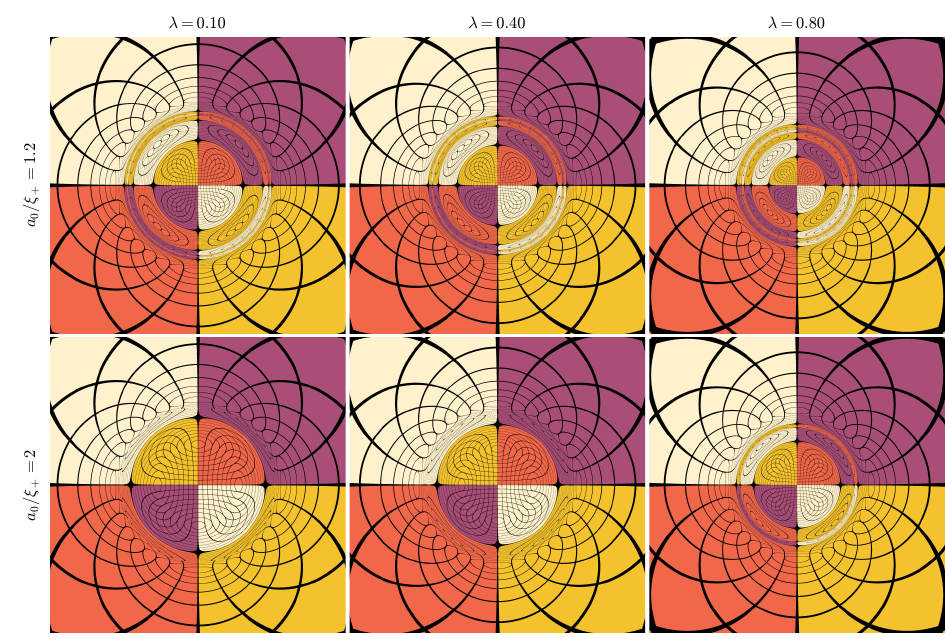}
    \caption{Celestial light sphere images of the symmetric magnetic thin-shell wormhole at fixed asymptotic mass $M$. The columns have $\lambda=0.10,0.40,0.80$, and the upper and lower rows have $a_0/\xi_+=1.2$ and $2$, respectively. An equatorial static ZAMO at $\xi_{\rm o}=200\mu$ views identical diagnostic spheres at $\xi_{\rm s}=1000\mu$ in both exteriors. All panels share a stereographic field of $6^\circ$ per axis, centered on the inward optical axis, with the launched-ray orientation of Eq.~\eqref{eq:magnetic_opt_camera_direction}. The latitude pattern and palette are those of Fig.~\ref{fig:exponential-minkowski-grid}; both sources use the same colors and contain no additional point source. Colors label source positions, while black curves are images of the source grid; neither represents disk intensity.}
    \label{fig:exponential-celestial-spheres}
    \end{minipage}
\end{figure*}

The upper row of Fig.~\ref{fig:exponential-celestial-spheres} contains progressively narrower annular replicas around the return-transmission boundary. This near-critical behavior is associated with the logarithmic strong-deflection regime of a retained nondegenerate unstable photon sphere~\cite{Bozza2002}. The retained maximum of the null potential allows near-critical rays to accumulate large orbital angles in Eq.~\eqref{eq:magnetic_opt_angle_integral}, repeatedly mapping the same source pattern onto the observer's sky. At finite angular resolution, the finest replicas merge into narrow bands. Moving the throat to the lower-row radius removes the bulk photon sphere and replaces its critical orbit by the throat-grazing branch of Eq.~\eqref{eq:magnetic_opt_threshold_branches}. The central pattern is consequently less densely replicated, although ordinary lensing still produces curved arcs and annular images. The critical curve separates endpoints in the two exteriors; with both spheres illuminated, it does not bound an absorbing dark region.

Increasing $\lambda$ through the three displayed values contracts the angular return-transmission boundary in both rows. Its calculated diameter decreases from $2.901^\circ$ to $2.310^\circ$ in the upper row, and from $3.150^\circ$ to $2.349^\circ$ in the lower row. These changes are measured on a common angular scale, rather than on panels rescaled by their individual photon impact parameters. For the retained light ring, the contraction follows the decreasing $\mathfrak b_{\rm ph}/\mu$ in Table~\ref{tab:magnetic-optical-scales}, together with the finite-observer conversion in Eq.~\eqref{eq:magnetic_opt_screen_impact}. For the excised light ring, both the lapse at the throat and its radius determine the boundary. A fixed value of $a_0/\xi_+$ therefore compares different physical throat radii as the magnetic geometry changes, rather than holding fixed $a_0/\mu$.

The lower right panel retains an additional annular replica despite the absence of a bulk photon sphere. The finite grazing limit of the source-angle sweep is approximately $5.361$, $5.679$ and $8.098\,\mathrm{rad}$ for the three columns, respectively. Only the last value exceeds $2\pi$, allowing another image of the aligned intersection of the two source-latitude families through Eq.~\eqref{eq:magnetic_opt_source_endpoint}. This finite extra winding does not restore the logarithmically divergent light-ring hierarchy. For the same $\lambda$, the outer parts of the two rows remain unchanged wherever the returning rays turn outside both candidate throats, because their two exterior angular integrals in Eq.~\eqref{eq:magnetic_opt_source_angles} are then identical. The altered inner replicas arise from the rays whose accessible radial domain changes at the junction.

The apparent exchange of sector colors relative to Fig.~\ref{fig:exponential-minkowski-grid} follows from Eq.~\eqref{eq:magnetic_opt_source_endpoint}. In the corresponding flat view, nonradial inward rays have $0<\varphi<\pi$, giving both tangential components a positive common sine factor. For $\pi<\varphi<2\pi$, this factor reverses both components at fixed screen azimuth. Golden yellow and ivory are opposite sectors, as are coral and wine. Representative peripheral rays in Fig.~\ref{fig:exponential-celestial-spheres} accumulate approximately $200^\circ$--$202^\circ$, selecting these opposite sectors under deflection with the same palette and signed camera convention. Successive sine-sign changes contribute to the alternating sectors in the inner annuli.

The source-latitude levels are symmetric about zero, so this sign reversal alone preserves black-band membership. Curve bending and annular replication follow the full source-angle map in Eq.~\eqref{eq:magnetic_opt_source_angles}. Thus the $10^\circ$ spacing on the source sphere is distinct from the lensed angular spacing. A flat view with the same $6^\circ$ field contains only the central zero-latitude cross and four sectors, whereas deflection brings images of higher latitude circles into the wormhole field. Independent ray controls in Minkowski spacetime recover the grid masks and sector colors obtained from Cartesian straight-ray intersections at the same observer and source radii.

The two sources have identical palettes, so a given sector color alone does not identify the exterior of origin. Source identity follows from the returning or transmitted ray branch. In particular, the ivory regions belong to the prescribed texture and are not measurements of enhanced brightness. The annular structure thus measures the lens mapping under the specified illumination, whereas the specific intensity of a moving disk is governed separately by Eq.~\eqref{eq:magnetic_opt_intensity_sum}.

\begin{figure}[tbp]
    \noindent\begin{minipage}{\columnwidth}
    \centering
    \includegraphics[width=\linewidth]{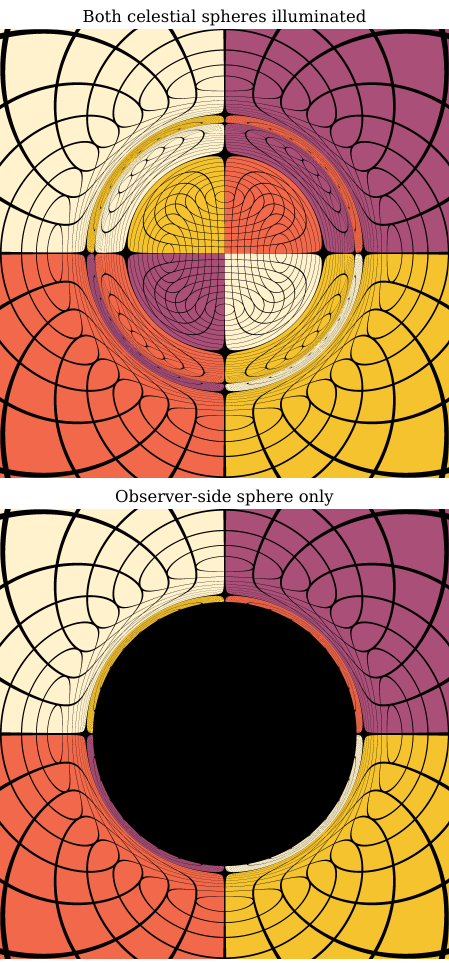}
    \caption{Celestial light sphere images for $\lambda=0.80$ and $a_0/\xi_+=2$, with a stereographic field of $4^\circ$ per axis. Both spheres are illuminated in the upper panel; only the observer-side sphere is illuminated in the lower panel. An equatorial static ZAMO at $\xi_{\rm o}=200\mu$ views sources at $\xi_{\rm s}=1000\mu$, with the grid and palette of Fig.~\ref{fig:exponential-minkowski-grid} and the camera orientation of Eq.~\eqref{eq:magnetic_opt_camera_direction}. The geodesic map is identical in both panels. The central dark region in the lower panel contains transmitted directions without an illuminated source. Colors identify source positions rather than disk intensity; no additional white source is present.}
    \label{fig:magnetic-celestial-source-sides}
    \end{minipage}
\end{figure}

Figure~\ref{fig:magnetic-celestial-source-sides} isolates the effect of the opposite-side source on a fixed lens map. Switching off that source removes the central colored replicas, while the surrounding arcs and sector boundaries coincide in the two panels. These surviving structures are returning images of the observer-side sphere, with source angles given by Eq.~\eqref{eq:magnetic_opt_source_angles}. The transmitted trajectories still cross the transparent junction and reach the other exterior; their dark endpoints result from the absence of illumination there. The change therefore distinguishes the two ray branches without changing their paths or introducing absorption at the shell.

The dark-region boundary has an angular diameter of approximately $2.349^\circ$, equal to the return-transmission diameter of the lower-right panel in Fig.~\ref{fig:exponential-celestial-spheres}. The narrower field resolves the same boundary at a larger fraction of the image width. For this throat, the seed photon sphere has been excised, so the boundary follows the throat-grazing branch of Eq.~\eqref{eq:magnetic_opt_threshold_branches}, rather than a retained bulk light ring. Its interpretation depends on the illumination: the same transmitted domain is filled by source images in the upper panel. Thus a dark central region in this comparison does not imply a horizon or loss of traversability.

An inward ray that crosses the throat passes an identical potential barrier in the other exterior and cannot turn back toward the junction. Accessible rays from the distant observer therefore have zero or one throat crossing. Celestial darkness does not by itself imply a dark disk pixel: disk radiation accumulates along supported intersections before the ray reaches its celestial endpoint.

Both throat choices remain below $\xi_{\rm I}$ for the geometries in Table~\ref{tab:magnetic-optical-scales}, so the same ISCO-truncated disk law is admissible in each case. For disk radiation, bilateral illumination adds nonnegative contributions,
\begin{equation}
 I_{\nu,\rm o}=I_{\nu,\rm o}^{(+)}+I_{\nu,\rm o}^{(-)},\qquad
 I_{\nu,\rm o}^{(\varsigma)}\geq0.
 \label{eq:magnetic_opt_side_sum}
\end{equation}
Consequently, switching on the opposite disk cannot lower the raw intensity of a fixed pixel. Its effect is confined to rays with supported opposite-side crossings and need not be spatially uniform. The common map in Eq.~\eqref{eq:magnetic_opt_display} preserves the ordering of positive intensities without turning source-sector colors into brightness measurements.

Inclination affects the rotating source through $\mathfrak b_\phi$ in Eq.~\eqref{eq:magnetic_opt_frequency_circular}. At the same emitting radius, positive $\mathfrak b_\phi$ enhances the frequency factor for the chosen positive rotation. This produces a Doppler contrast without breaking the reflection symmetry of the geometry. On the disk axis, $\mathfrak b_\phi=0$ and the image is azimuthally symmetric; the transverse-motion factor in the numerator remains and cannot be replaced by the redshift of static gas. Actual intensity contrasts also depend on the crossing radii and source support. Finally, at fixed $(\lambda,a_0)$, the three surface equations of state share the same stationary optical map. Their distinct radial restoring curvatures in Sec.~\ref{sec:constitutive} constrain the persistence of that equilibrium, rather than providing additional stationary image parameters.

\begin{figure*}[t]
    \noindent\begin{minipage}{\textwidth}
    \centering
    \includegraphics[width=\linewidth]{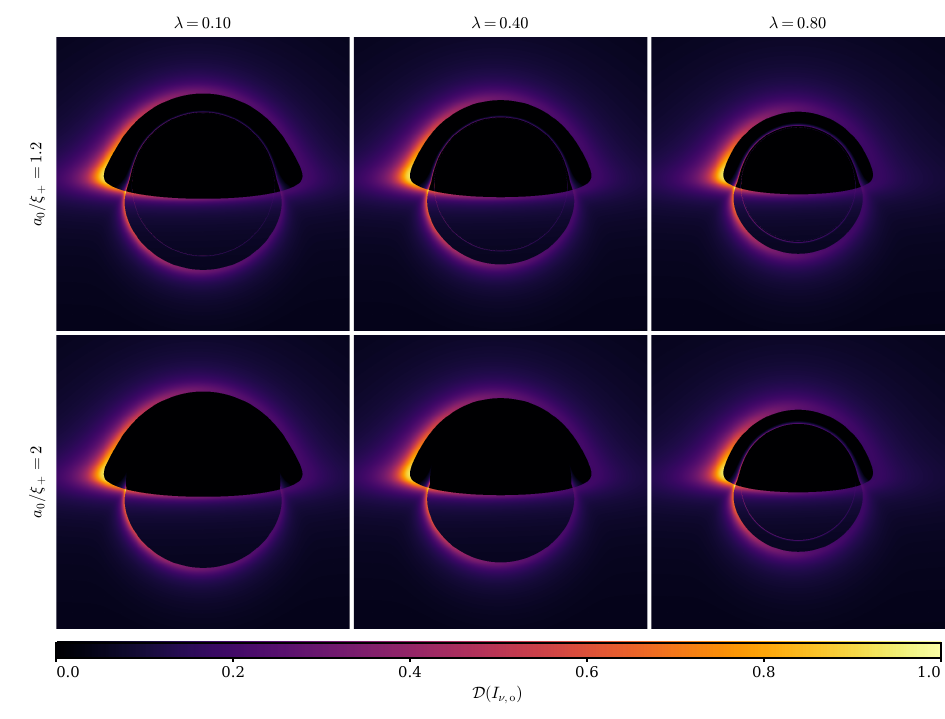}
    \caption{Observed specific-intensity images of the moving thin disk at $\iota=80^\circ$. The columns have $\lambda=0.10,0.40,0.80$, and the upper and lower rows have $a_0/\xi_+=1.2$ and $2$, respectively, at fixed asymptotic mass $M$. The static ZAMO is at $\xi_{\rm o}=200\mu$, with a stereographic field of $6^\circ$ per axis and the disk-image orientation specified in Sec.~\ref{subsec:optical_imaging_method}. Only the observer-side disk emits. Circular gas follows Eq.~\eqref{eq:magnetic_opt_circular_velocity}, with the ISCO-truncated GLM profile in Eq.~\eqref{eq:magnetic_opt_emissivity} and outer edge $1000\mu$. All six panels use the same logarithmic intensity map, Eq.~\eqref{eq:magnetic_opt_display}, whose reference is their common sampled maximum. Colors encode transported specific intensity, including the cubic frequency factor and all supported disk crossings; no celestial texture is superposed.}
    \label{fig:magnetic-inclined-disk}
    \end{minipage}
\end{figure*}

The nearly edge-on view in Fig.~\ref{fig:magnetic-inclined-disk} separates a narrow, bright disk image from curved lensed structures extending above and below it. The left side is brighter for the chosen rotation, consistently with the sign of $\mathfrak b_\phi$ in Eq.~\eqref{eq:magnetic_opt_screen_impact} and its frequency weighting in Eq.~\eqref{eq:magnetic_opt_frequency_circular}. The luminous arcs are images of emitting matter at different ray crossings, rather than material rings placed at the photon sphere. Their large apparent vertical extent is compatible with an infinitesimally thin equatorial source: the ray paths are deflected out of the observer's direct view of that plane.

For the upper-row throats, the observer-side disk image coincides with that of the corresponding magnetic black hole seed under the same observer, emission and screen prescriptions. Table~\ref{tab:magnetic-optical-scales} places these throats below the seed photon sphere and the disk inner edge. Every returning ray turns outside the photon sphere and therefore traverses the same exterior in both objects. An inward ray that would be captured by the seed instead crosses the throat, but its observer-side emitting intersections are already complete before the trajectories separate: neither model emits below the ISCO, and the opposite wormhole disk is switched off. Consequently, the crossing sum and frequency factors in Eq.~\eqref{eq:magnetic_opt_intensity_sum} agree. This equivalence persists at inclined viewing angles; it is conditional on the specified transparent, one-sided, ISCO-truncated source and does not equate the global geometries or their causal structures.

Moving the throat to the lower row radius preserves the bright primary structure but suppresses part of the fainter inner image. The dominant left enhancement and the broad surrounding emission remain very similar within each column, whereas the narrow interior arcs are removed or weakened. At this radius the bulk photon sphere is excised. Some rays that returned in the smaller throat geometry now pass into the non-emitting exterior before completing their observer-side return crossings. Rays that turn outside both throats, and emitting intersections reached before the altered path segment, remain unchanged. The difference between rows is therefore concentrated in selected lensed contributions rather than a uniform dimming or a displacement of the entire disk image. In particular, the most conspicuous row differences concern faint structure and are much smaller than the contrast between the bright approaching side and the rest of the disk.

Increasing $\lambda$ contracts the overall image on the common angular screen. The inner dark region, the bright band and the surrounding arcs occupy smaller angular scales, in agreement with the decreasing ISCO and critical impact scales in Table~\ref{tab:magnetic-optical-scales}. At fixed $a_0/\xi_+$, both the physical throat radius in mass units and the exterior geometry vary, so this column comparison does not isolate a displacement of the throat in a fixed metric. The upper-row sampled intensity peaks are about $2.9$, $3.0$ and $3.1$ in units of $I_*$, respectively; these finite-resolution maxima are illustrative rather than convergence-certified peak values. Thus the contraction does not imply a lower maximum intensity; the common color scale retains the modest increase of the approaching side peak in this scan.

The lower-right panel remains particularly close to its upper-row counterpart and retains a faint inner arc that is strongly suppressed in the first two lower-row panels. For $\lambda=0.80$, the larger throat lies approximately $14\%$ above the seed photon-sphere radius, compared with $33\%$ and $31\%$ for $\lambda=0.10$ and $0.40$. Surviving returning trajectories can therefore approach the seed photon region more closely in the last column, consistently with the additional resolved inner structure. The bulk photon sphere itself is nevertheless absent from every lower-row spacetime. The surviving finite lensed arcs do not establish the divergent hierarchy associated with a retained unstable circular null orbit. Under this one-sided emission prescription, the similarity of the brightest structures alone consequently provides little discrimination between the two throat radii; their differences are more apparent in the faint inner branches.

\begin{figure}[tbp]
    \noindent\begin{minipage}{\columnwidth}
    \centering
    \includegraphics[width=\linewidth]{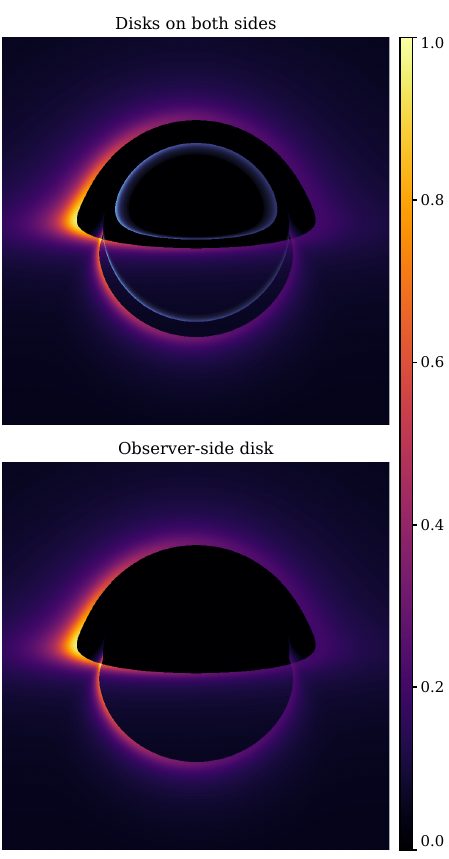}
    \caption{Bilateral and observer-side thin-disk emission for $\lambda=0.40$, $a_0/\xi_+=2$ and $\iota=80^\circ$. Identical disks emit in both exteriors in the upper panel; only the observer-side disk emits below. The static ZAMO is at $\xi_{\rm o}=200\mu$, with a $6^\circ$ stereographic field per axis and the disk-image orientation of Sec.~\ref{subsec:optical_imaging_method}. Both disks have the circular flow and ISCO-truncated GLM profile of Fig.~\ref{fig:magnetic-inclined-disk}. The vertical bar shows $\mathcal{D}(I_{\nu,\rm o})$ from Eq.~\eqref{eq:magnetic_opt_display}, using a common normalization for both panels. Warm and cool hues identify observer-side and opposite side emission, respectively; mixed colors mark overlapping contributions.}
    \label{fig:magnetic-bilateral-disk}
    \end{minipage}
\end{figure}

Illuminating the second disk in Fig.~\ref{fig:magnetic-bilateral-disk} adds a closed inner arc above the bright equatorial band and a curved contribution below it. These structures occupy part of the region that is dark or faint when only the observer-side disk emits. The broad warm crescent and its strong left-side enhancement remain similar in both panels because their observer-side contribution is unchanged. In particular, the lower panel contains the same raw specific intensity map as the lower middle panel of Fig.~\ref{fig:magnetic-inclined-disk}. Its display uses the common maximum of this pair, rather than the maximum of that six panel scan, so differences in displayed brightness between the two figures do not imply a change of the observer-side source.

The additional radiation is carried by rays that cross the transparent throat and intersect the opposite disk within its emitting support. Returning rays that remain in the observer's exterior receive no contribution from that disk. Equation~\eqref{eq:magnetic_opt_side_sum} therefore accounts for the localized increase without changing the trajectories or the existing emission. The remaining central darkness in the bilateral image corresponds to directions without a supported emitting intersection, even though the spacetime has no retained horizon. A central dark region alone thus does not identify capture by a black hole or an opaque throat. Unlike the extended celestial illumination in Fig.~\ref{fig:magnetic-celestial-source-sides}, the thin disks occupy only their equatorial, ISCO-truncated domains and need not illuminate every transmitted direction.

The cool rendering identifies the exterior of emission and is not a physical blueshift. The two source components retain the frequency transfer of Eq.~\eqref{eq:magnetic_opt_frequency_circular}; the static symmetric junction itself supplies no additional frequency jump. Their raw intensities are summed before the common logarithmic transformation, while the displayed hue is mixed in proportion to their individual contributions. The warm color bar consequently specifies the nonlinear intensity coordinate, but cannot by itself decode a mixed or cool hue into a unique intensity.

At the available image sampling, the opposite disk contributes to about $4\%$ of the pixels and increases the unweighted sum of specific intensities over the fixed screen by about $8\%$. These rounded values describe that numerical realization and are not assigned sub-percent accuracy. Its effect is therefore concentrated in the inner image rather than distributed uniformly over the field. These percentages concern the resolved pixels and their specific-intensity sums within the stated angular window; they are not a bolometric luminosity or a fraction of the total emission over the sky. The combination of nearly unchanged bright outer structures with additional inner arcs isolates the effect of the source distribution at fixed geometry.

\section{Conclusion}\label{sec:conclusion}
The surface material and the illumination of the two exteriors control distinct properties of the symmetric wormholes obtained from the exponential magnetic black-hole geometry. The junction fixes the equilibrium density and pressure, whereas the response to a radial displacement depends on the chosen pressure law. At the same geometric equilibrium, different surface fluids can therefore have different restoring behavior while producing the same stationary optical map. This separation allows the local radial response of a throat to be assessed independently of the source assumptions underlying its image.

For the four magnetic configurations examined, explicit radius dependence in the variable phantom and variable Chaplygin laws admits locally stable radial sectors, while the calibrated linear barotropic law remains unstable throughout the displayed domain. In the phantom family, a larger positive radial exponent strengthens the restoring contribution and shifts the stability boundary inward. The inverse-density response of the Chaplygin law instead cancels the inverse-lapse term responsible for the divergent perturbative curvature near a simple horizon. Stable sectors can consequently extend close to that horizon. The strong overlap of the Chaplygin curves at larger radii follows from cancellation of the leading magnetic-charge contribution, rather than exact independence from the magnetic geometry. These statements concern linear radial modes with each material calibration held constant. They do not establish stability against nonspherical, electromagnetic or coupled bulk-shell perturbations, and the phenomenological equations of state do not supply a microscopic realization of the negative density layer.

The stable sectors retain the negative surface density required by the symmetric junction. Positive pressure allows the tangential null and intrinsic strong energy conditions to hold over an inner radial interval, while the weak and dominant energy conditions fail. The shell-density minimum and saturation of the tangential null condition have the same geometric location. At fixed asymptotic mass, increasing the dimensionless magnetic shape parameter $\lambda$ moves this location and the outer seed horizon inward along the sampled family. Taking the throat closer to a nonextremal horizon reduces the magnitude of its density but makes its pressure singular, so this limit does not yield a shell with vanishing stresses. For reversible local equilibrium at fixed composition, the transparent dynamic shell evolves isentropically even in the nonvacuum magnetic bulk. Its exact acceleration--stress relation supplies a geometric temperature scale, while the material temperature and thermodynamic stability require an additional thermodynamic completion. Under a fixed-coefficient extension of the three surface laws, the Maxwell relation fixes their relative isentropic temperature evolution. At the photon-sphere equilibrium, the two variable fluids share the same area stiffness and radial curvature but have opposite thermal responses.

For independent light minimally coupled to the gravitational metric, a transparent throat below the seed photon sphere can reproduce the seed black-hole disk image, including an inclined view, when emission is restricted to the observer's exterior and begins at the innermost stable circular orbit. The retained exterior then supplies the same emitting intersections for returning rays. Placing the throat above the photon sphere removes some return contributions and alters the faint inner branches more strongly than the dominant bright structure. The persistence of finite lensed arcs in this case is compatible with excision of the bulk circular null orbit. Across the imaged magnetic configurations, increasing $\lambda$ contracts the angular structures without requiring a lower peak specific intensity.

Illumination determines which transmitted directions become visible. The celestial-source comparisons isolate this dependence, and the rotating-disk images show additional inner arcs when the opposite exterior contains an emitting disk. The observer-side contribution is preserved while the second source adds radiation along supported cross-throat paths. A central dark region can nevertheless persist because a truncated thin disk does not illuminate every such direction. The image therefore constrains the combination of retained geometry and emitting support; its dark area alone cannot identify an event horizon. This statement applies to the independent probe and idealized surface emission used here. It neither establishes degeneracy for a self-consistent accretion flow nor replaces the characteristic ray treatment required for the supporting nonlinear electromagnetic field.

The new geometric connections sharpen the relation between the two parts of the calculation. Photon-sphere equilibrium is simultaneously a stationary-density minimum, tangential NEC saturation point, and a radius-independent barotropic stability obstruction. Explicit radius pressure response changes the last property while leaving the first two intact. The imaged radii admit the material dependent classifications in Table~\ref{tab:imaged_stability_thresholds}, including stable radial realizations of the one-sided black-hole image degeneracy. The exact radial ANEC result in Eq.~\eqref{eq:radial_anec_exact} also shows why vanishing proper shell energy in a near horizon limit does not eliminate averaged null energy violation. These statements preserve the distinction between local radial support, global energy condition violation, and the illumination dependence of the stationary optical map.

A joint parameter study of shell oscillation frequencies and resolved inner-image structure would extend the equilibrium classification already made here. Evaluating both quantities for the same equilibria could determine how dynamical information complements imaging when separating the material response from the magnetic geometry and source distribution.

\begin{acknowledgments}
Edson Otoniel would like to thank Fundação Cearense de Apoio ao Desenvolvimento Científico e Tecnológico through grant BP6-0241-00335.01.00/25. Francisco S. N. Lobo acknowledges funding from the Fundação para a Ciência e a Tecnologia (FCT) through the research grant UID/04434/2025 and the FCT Scientific Employment Stimulus contract with reference CEECINST/00032/2018.
\end{acknowledgments}

\bibliographystyle{apsrev4-2}
\bibliography{ref}
\end{document}